\documentclass[%
 reprint,
superscriptaddress,
 amsmath,amssymb,
 aps,
]{revtex4-2}

\usepackage{graphicx}% Include figure files
\usepackage{dcolumn}% Align table columns on decimal point
\usepackage{bm}% bold math
\usepackage{dcolumn}
\usepackage{array,multirow}
\usepackage{hyperref}% add hypertext capabilities
\usepackage{xcolor}
\begin{document}

\preprint{APS/123-QED}

\title{Exciton landscape in van der Waals heterostructures}% Force line breaks with \\

\author{Joakim Hagel}
  \email{joakim.hagel@chalmers.se}
  \affiliation{%
Department of Physics, Chalmers University of Technology, 412 96 Gothenburg, Sweden\\
}%
\author{Samuel Brem}%
\affiliation{%
 Department of Physics, Philipps University of Marburg, 35037 Marburg, Germany\\
}%
 \author{Christopher Linderälv}%
 \affiliation{%
Department of Physics, Chalmers University of Technology, 412 96 Gothenburg, Sweden\\
}%
 \author{Paul Erhart}%
 \affiliation{%
Department of Physics, Chalmers University of Technology, 412 96 Gothenburg, Sweden\\
}%
  \author{Ermin Malic}%
  \affiliation{%
 Department of Physics, Philipps University of Marburg, 35037 Marburg, Germany\\
}%
\affiliation{%
Department of Physics, Chalmers University of Technology, 412 96 Gothenburg, Sweden\\
}%

\date{\today}

\begin{abstract}
Van der Waals heterostructures consisting of vertically stacked transition metal dichalcogenides (TMDs) exhibit a rich landscape of bright and dark intra- and interlayer excitons. In spite of a growing literature in this field of research, the type of excitons dominating optical spectra in different van der Waals heterostructures has not yet been well established. The spectral position of exciton states depends strongly on the strength of hybridization and energy renormalization due to the periodic moir\'e potential. Combining exciton density matrix formalism and density functional theory, we shed light on the exciton landscape in TMD homo- and heterobilayers at different stackings. This allows us to identify on a microscopic footing the energetically lowest lying exciton state for each material and stacking. Furthermore, we disentangle the contribution of hybridization and layer polarization-induced alignment shifts of dark and bright excitons in photoluminescence spectra. By revealing the exciton landscape in van der Waals heterostructures, our work provides the basis for further studies of the optical, dynamical and transport properties of this technologically promising class of nanomaterials. 
\end{abstract}

\maketitle

\section{Introduction}
The emergence of atomically thin semiconductors has opened up a new research venue \cite{ares2021recent}. In particular, vertically stacked van der Waals heterostructures have recently gained much attention \cite{geim2013van,jin2018ultrafast,liao2019van,Merkl2019}, displaying intriguing fundamental properties, such as the appearance of moir\'e patterns and correlated states \cite{cao2018unconventional,yu2017moire,tong2017topological,sung2020broken,alexeev2019resonantly,tran2019evidence,brem2020tunable}. The transition metal dichalcogenides (TMDs) are particular promising representatives of this class of materials. Due to their strong Coulomb interaction, they harbor stable exciton features dominating optical, dynamical and transport properties of these materials even at room temperature \cite{wang2018colloquium,mueller2018exciton,deilmann2019finite,Berghauser2018}. By vertically stacking two TMD monolayers on top of each other into a so-called van der Waals heterostructure, long-lived, spatially separated interlayer excitons can form \cite{kunstmann2018momentum,rivera2018interlayer,jiang2021interlayer,gillen2018interlayer}. Through an interlayer wave function overlap, these excitons can efficiently hybridize and form new hybrid excitons \cite{wilson2017determination,Brem19b}, as shown in \autoref{fig:Schematic}. Previous studies on interlayer excitons have been conducted, focusing mostly on their optical signatures \cite{gillen2018interlayer,deilmann2019finite,enalim2019restoring,cao2019phonon,rivera2015observation,zheng2015coupling,jones2014spin}. However, the crucial question about the spectral position and ordering of bright and dark exciton states is still debated.

In this work, we combine the density matrix formalism with density functional theory (DFT) calculations to reveal the exciton landscape for untwisted vertically stacked TMD homo- and heterobilayers. In particular, we identify the lowest lying exciton state for different TMD homo- and heterobilayers at different stackings. Furthermore, we disentangle the spectral shifts of different exciton types (\autoref{fig:Schematic}c) stemming from hybridization and energy renormalization due to a charge polarization of the two layers (resulting in a periodic moir\'e potential). We find a strong impact of hybridization on momentum-dark excitons, which can turn them into the energetically lowest states and the material into an indirect semiconductor. We also analyze the optical footprint of the exciton landscape by calculating photoluminescence (PL) spectra and taking into account direct and phonon-assisted exciton recombination. The latter allows an indirect visualization of dark excitons via the emergence of phonon sidebands \cite{brem2020phonon}. This allows us to identify the microscopic origin of resonances appearing in PL spectra of different TMD homo- and heterobilayers at different stackings. 

\section{Theory}
\begin{figure}[t!]
\includegraphics[width=7cm]{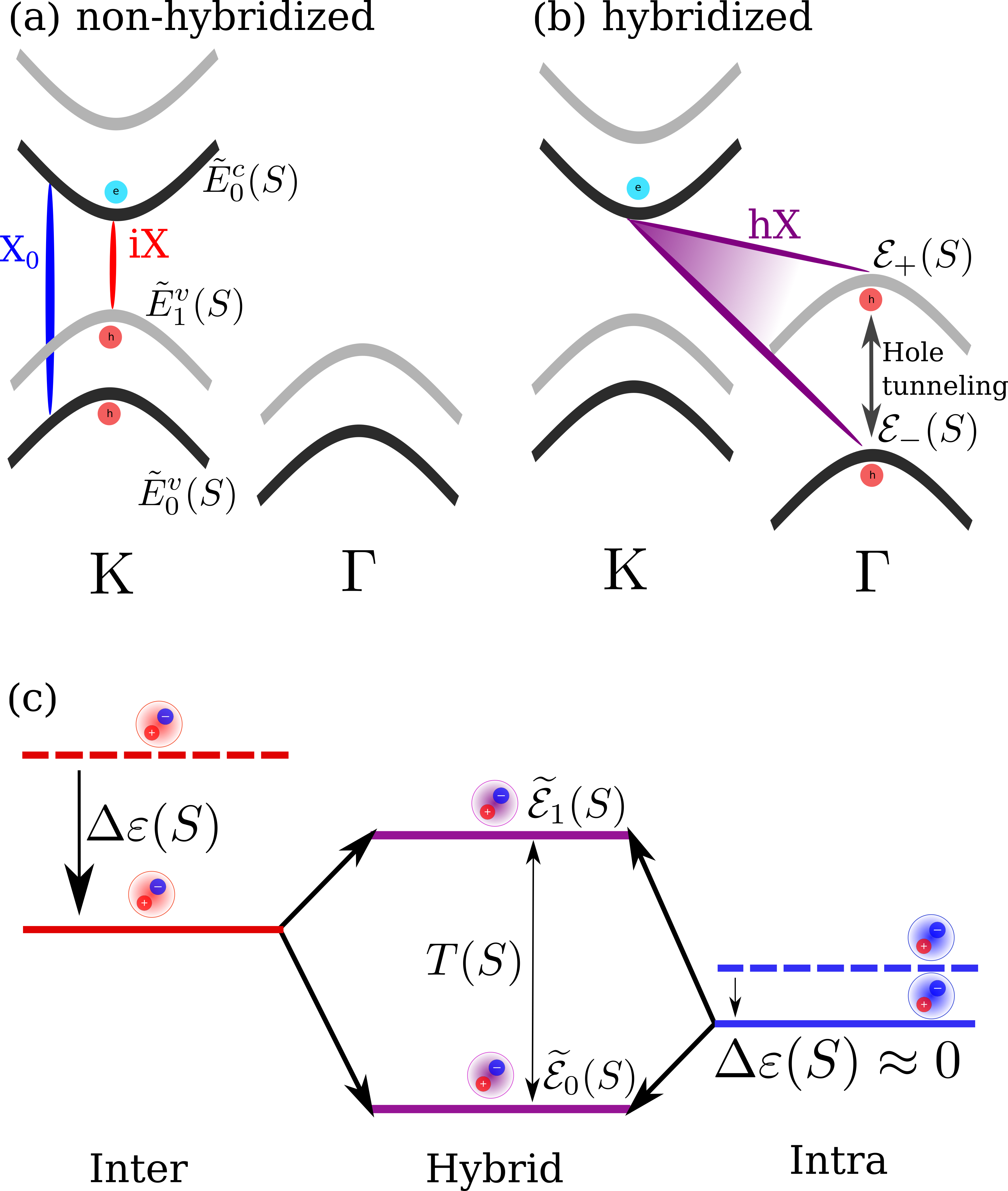}
\caption{\label{fig:Schematic} $\bm{(a)}$ Schematic of the electronic band structure in MoS$_2$(black)-WS$_2$(gray) heterostructure (a) before and (b) after hybridization. The intralayer (X$_0$) and interlayer (iX) excitons are marked in blue and red, respectively. The strong tunneling of holes around the $\Gamma$ point results in a pronounced hybrid exciton state (hX, purple line). $\bm{(c)}$ Schematic for the formation of hybrid excitons. The dashed lines are the unperturbed exciton energies that become shifted by $\Delta\varepsilon(S)$ due to the layer polarization. Interlayer hybridization results in hybrid exciton states denoted by $\widetilde{\mathcal{E}}_{\eta}(S)$.}
\end{figure}

We exploit the exciton density matrix formalism \cite{ivanov1993self,katsch2018theory,kira2011semiconductor} with input from DFT to construct a material specific fully microscopic framework for modeling the exciton energy landscape in van der Waals heterostructures. The goal is to disentangle the microscopic contributions from hybridization and layer polarization-induced alignment shifts, to the final energies of hybrid excitons, as illustrated in \autoref{fig:Schematic}.

We start by formulating a many-particle Hamilton operator in second quantization. The interaction-free electronic Hamiltonian for vertically stacked TMDs consists of three parts $H=H_0+H_{ren}+H_T$, where $H_0$ is the stacking-independent kinetic energy. The renormalization Hamiltonian $H_{ren}$ describes the stacking-dependent layer polarization-induced alignment shift \cite{brem2020tunable,D0NR02160A}. In the following, we refer to this as the alignment shift. Finally, $H_T$ is the hybridization Hamiltonian taking into account the overlapping electronic wavefunctions giving rise to  hybrid exciton states. In the basis of monolayer eigenstates, the bilayer Hamiltonian thus reads
\begin{equation}\label{eq:Hamiltonian}
    H=\sum_{\alpha l\bm{k}}\tilde{E}^{\alpha}_{l\bm{k}}(S)\, a^{\dagger}_{\alpha l\bm{k}}a_{\alpha l\bm{k}}+\sum_{\substack{\alpha \bm{k}\\ l\neq l^{\prime}}}T^{\alpha}_{ll^{\prime}}(S)\,a^{\dagger}_{\alpha l\bm{k}}a_{\alpha l^{\prime}\bm{k}},
\end{equation}
where $l$/$l^{\prime}$ are layer indices while $\alpha=(\lambda,\xi)$ is a compound index with $\xi$ denoting the valley and $\lambda=(c,v)$ the conduction and the valence band, respectively. The goal of this framework is to explicitly take into account dark intervalley exciton states \cite{malic2018dark}. Here, we have introduced the electronic creation (annihilation) operators $a^{\dagger}(a)$. Furthermore, $\tilde{E}^{\alpha}_{l\bm{k}}(S)=E^{\alpha }_{l\bm{k}}+\Delta\varepsilon^\lambda_l(S)$ includes the stacking-independent kinetic energy of the monolayers ($E^{\alpha }_{l\bm{k}}$) and the alignment shift ($\Delta\varepsilon^\lambda_l(S)$). 

Finally, $T^{\alpha}_{ll^{\prime}}(S)$ is the tunneling matrix element, where the stacking dependence ($S$) results from the varying interlayer distance and orbital symmetry of the Bloch functions involved. For the K point, the conduction and valence band Bloch waves are composed of d-orbitals \cite{D0NR02160A,wang2017interlayer,cappelluti2013tight}. Using the angular symmetry of d-orbitals, the tunneling around the three equivalent K points in an untwisted structure can be written as \cite{D0NR02160A} 
\begin{equation}\label{eq:TmatrixElement}
    T^{\alpha}_{ll^{\prime}}(S)=\frac{1}{3}\sum_{n=0}^{2}t^{\alpha}(S)\,e^{i\tau(\text{C}^{(n)}_3\bm{K}-\bm{K})\cdot \bm{D}(S) }, \end{equation}
where $\text{C}^{(n)}_3$ denotes the three-fold rotation operator and $\bm{D}(S)$ the stacking dependent lateral displacement between the layers. Furthermore, $t^{\alpha}(S)$ describes the tunneling strength and $\tau$ is a prefactor which equals $1$ for the K and $-1$ for the K$^{\prime}$ point. The factor $1/3$ reflects that the three equivalent K points have the same tunneling strength. For H-type structures, this expression obtains the additional phase $e^{i(-1)^{l^{\prime}}2\pi /3}$ for the valance band \cite{D0NR02160A}. 

Due to symmetry, tunneling can only occur for certain stacking configurations. To demonstrate this effect, we first consider tunneling around the K point for R-type structures (schematically shown in \autoref{fig:bandstruct}c) and apply the corresponding lateral displacement vectors $\bm{D}(S)$ for each high-symmetry stacking. We find that the tunneling term vanishes for $R_h^M$ and $R_h^X$ around the K point \cite{ruiz2019interlayer}. For an H-type structure one of the two layers is inverted (rotated by $180^{\circ}$) relative to the R-structure. Here, the tunneling for both electrons and holes vanishes at $H^M_h$. In addition, hole (electron) tunneling does not occur at $H^X_h$ ($H_h^h$). This means that we only have a non-zero tunneling matrix element at the K point for $R_h^h$, $H_h^{X}$ (electron tunneling) and $H_h^{h}$ (hole tunneling) stacking.
In these cases, \autoref{eq:TmatrixElement} simply becomes $T^{\alpha}_{ll^{\prime}}(S)=t^{\alpha}(S)$. 

For tunneling around the $\Gamma$ and $\Lambda$ points, there are no equivalent points within the first Brillouin zone (BZ) and thus the tunneling matrix element is only given by the tunneling strength $T^{\alpha}_{ll^{\prime}}(S)=t^{\alpha}(S)$ \cite{D0NR02160A}. The latter can be extracted from DFT calculations of the bilayer band structure by considering the Hamiltonian in \autoref{eq:Hamiltonian} as a $2\times2$ matrix with respect to the layer index $l$/$l^{\prime}$. Here, the diagonal components are given by $\tilde{E}^{\alpha}_{l\bm{k}}(S)$ and the off-diagonal terms correspond to the tunneling $T^{\alpha}_{ll^{\prime}}(S)$. The eigenvalues of this matrix are given by the avoided-crossing formula
\begin{equation}\label{eq:crossing}
  \mathcal{E}^{\alpha}_{\pm ,\bm{k}}(S)=\frac{1}{2}\sum_{l=1}^2\widetilde{E}^{\alpha}_{l\bm{k}}(S)\pm\frac{1}{2}\sqrt{\widetilde{\Delta}^{\alpha}_{\bm{k}}(S)^2+4|t^{\alpha}(S)|^2},
\end{equation}
where $\widetilde{\Delta}^{\alpha}_{\bm{k}}(S)=\widetilde{E}^{\alpha}_{1\bm{k}}(S)-\widetilde{E}^{\alpha}_{2\bm{k}}(S)$ is the spectral difference between the monolayer energies as extracted from DFT calculations shifted by the layer polarization-induced alignment potential, sometimes referred to as the ferroelectric potential \cite{Bandenergylandscapes_Fal'ko} (\autoref{fig:Schematic}a). Furthermore, $|t^{\alpha}(S)|$ is the tunneling strength (see \autoref{eq:TmatrixElement}) and $\mathcal{E}^{\alpha}_{\pm ,\bm{k}}(S)$ are the hybrid energies corresponding to the bilayer eigenenergies extracted from DFT calculations. As illustrated in \autoref{fig:Schematic}b, the hybridization is particularly prominent at the $\Gamma$ point due to the strong tunneling probability of holes. In contrast to the bands at the K point, which are mostly composed of d-orbitals localized around the metal atom, the bands at the $\Gamma$ point also have a significant contribution from chalcogen atoms, thus considerably increasing the interlayer overlap \cite{cappelluti2013tight,roldan2014electronic,Brem19b}. 
Exploiting \autoref{eq:crossing}, the tunneling strength can be calculated for each band, valley, and stacking, yielding
\begin{equation}\label{eq:Tunnelling}
  |t^{\alpha}(S)|=\frac{1}{2}\sqrt{(\Delta\mathcal{E}^{\alpha}_{\bm{k}}(S))^2-\widetilde{\Delta}(S)^2},  \end{equation}
where $\Delta\mathcal{E}^{\alpha}_{\bm{k}}(S)=\mathcal{E}^{\alpha}_{+ ,\bm{k}}(S)-\mathcal{E}^{\alpha}_{- ,\bm{k}}(S)$ is the spectral difference between the hybridized electronic states.

To obtain a microscopic picture of the energy landscape in van der Waals heterostructure, we performed DFT calculations. First, we relax the investigated TMD homo- and heterobilayers using VASP \cite{KreHaf93, KreFur96, KreFur96_2}. The vdW-DF-cx functional \cite{BerHyl14} is used to account for the van der Waals interactions of the bilayers. The plane wave energy cutoff is 500~eV and the BZ is sampled using a $18\times 18\times 1$ $\boldsymbol{k}$ point mesh. The electron density of the relaxed structures is subsequently evaluated using the local density approximation in GPAW \cite{MorHanJac05, EnkRosMor10} with a grid spacing of 0.15~{\AA}. The alignment shift $\Delta\varepsilon^\lambda_l(S)$ stems from a charge transfer induced dipole moment \cite{tong2020interferences}. Its origin, the electrostatic potential is calculated by solving the Poisson equation. The alignment shifts can then be found within the project augmented wave formalism (Appendix A). 
Exploiting the computed monolayer energies ($E^{\alpha }_{l\bm{k}}$), alignment shifts  ($\Delta\varepsilon^\lambda_l(S)$) and the hybrid bilayer energies ($\mathcal{E}^{\alpha}_{\pm ,\bm{k}}(S)$), the tunneling strength can be calculated from \autoref{eq:Tunnelling}. The values are summarized for several homo- and heterobilayers in the appendices. 

Having obtained both the alignment shifts and the tunneling-induced hybridization, we investigate now the impact of these microscopic processes on the final exciton energy. To achieve this goal, we transform the Hamiltonian in \autoref{eq:Hamiltonian} into an exciton basis resulting in  \cite{katsch2018theory,brem2020tunable,D0NR02160A},
\begin{equation}\label{eq:ExcitonHamiltonian}
\begin{split}
    H=\sum_{\substack{\xi\bm{Q}\\LL^{\prime}}} (E^{\xi S}_{L\bm{Q}}X^{\xi \dagger}_{L  \bm{Q}}X^{\xi }_{L  \bm{Q}}\delta_{LL^{\prime}}+T_{LL^{\prime}}^{\xi}(S)X^{\xi\dagger}_{L\bm{Q}}X^{\xi}_{L^{\prime}   \bm{Q}}).
        \end{split}
\end{equation}
Here, $L=(l_e,l_h)$ and $\xi=(\xi_e,\xi_h)$ are compound layer and valley indices, the first labeling the two intra- and interlayer excitons respectively and the second labeling the valley configuration. The index $S$ denotes the stacking and $\bm{Q}$ the center-of-mass momentum. Furthermore, $X^{\dagger}_{L \xi  \bm{Q}}(X_{L \xi  \bm{Q}})$ is the exciton creation (annihilation) operator and $T_{LL^{\prime}}^{\xi}(S)$ is the exciton tunneling matrix element, which takes into account the tunneling strength and exciton form factors. The energy $E^{\xi S}_{L \bm{Q}}$ is the exciton energy associated with the binding energy, the stacking-independent monolayer band edges and the alignment shift. The completely stacking-independent part of this energy, i.e. without considering the alignment shifts, yields the unperturbed exciton energies (the dashed lines in \autoref{fig:Schematic}c). The interlayer and intralayer exciton binding energies are microscopically obtained by solving the Wannier equation , where the strong Coulomb interaction between electrons and holes is taken into account. Furthermore, the screening of the Coulomb potential is modeled with the generalized Keldysh screening \cite{ovesen2019interlayer}. The change in interlayer distance does in principle affect the screening, but since the variation in distance is small, this effect has an negligible impact on the binding energy.

The four different intra- and interlayer exciton states couple to each other through hybridization and split into four hybrid exciton states. The final hybrid exciton operator can be written in terms of a linear combination of the composing intra- and interlayer excitons i.e. $Y_{\xi\eta\bm{Q}}^{\dagger}=\sum_L\mathcal{C}^{\xi\eta}_{L}(\bm{Q})^*X_{\xi L \bm{Q}}^{\dagger}$, where $\eta$ is the hybrid exciton quantum number. The mixing coefficient $\mathcal{C}^{\xi\eta}_{L}(\bm{Q})$ fulfills the eigenvalue problem
\begin{equation}\label{eq:EigenVal}
    \widetilde{E}_{L\bm{Q}}^{\xi}(S)\mathcal{C}^{\xi\eta}_{L\bm{Q}}+\sum_{L^{\prime}}\widetilde{T}^{\xi}_{LL^{\prime}}(S)\mathcal{C}^{\xi\eta}_{L^{\prime}\bm{Q}}=\widetilde{\mathcal{E}}^{\xi}_{\eta \bm{Q}}(S)\mathcal{C}^{\xi\eta}_{L\bm{Q}}.
\end{equation}
Here, $\widetilde{\mathcal{E}}^{\xi}_{\eta \bm{Q}}(S)$ denotes the final hybrid exciton energies. Equation (\ref{eq:EigenVal}) provides microscopic access to the exciton landscape in different TMD homo- and heterobilayers, allowing us in particular to identify the energetically lowest exciton state for different stackings. 

\section{Results}
\subsection{Exciton energy landscape}
The final hybrid exciton energies are obtained by solving the eigenvalue problem in \autoref{eq:EigenVal}. We use the calculated tunneling strengths and alignment shifts, where the effective masses and valley separations are described in a monolayer basis \cite{Korm_nyos_2015}. This also yields the respective degree of hybridization for each exciton valley (K-K, K-K$^{\prime}$, K-$\Lambda$, $\Gamma$-K, $\Gamma$-K$^{\prime}$, $\Gamma$-$\Lambda$). As an exemplary material, we first investigate the untwisted MoS$_2$-WS$_2$ heterostructure on a SiO$_2$ substrate. The latter is taken into account via a substrate screening of the Coulomb potential.

\begin{figure}[t!]
\hspace*{-0.2cm}  
\includegraphics[width=9cm]{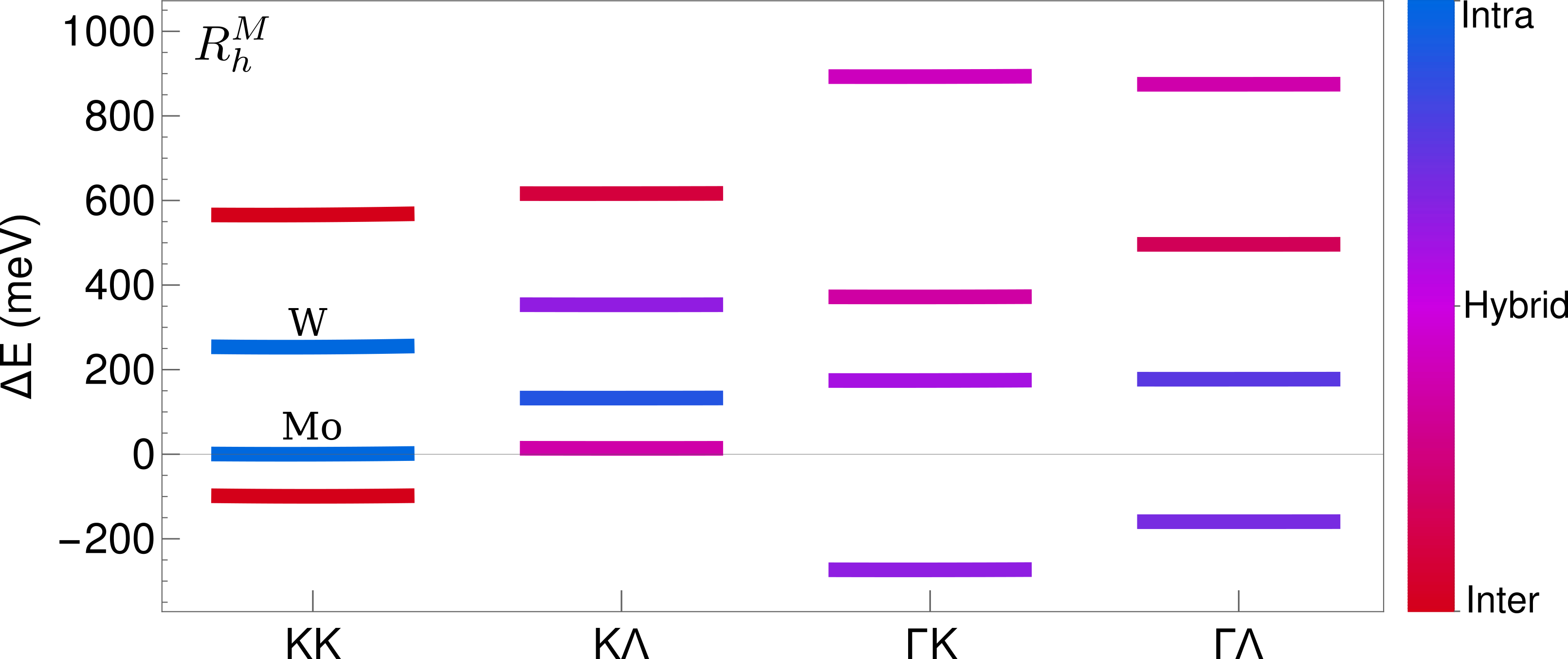}
\caption{\label{fig:ordering} Exciton band structure for all layer configurations at different valleys with $\bm{R_h^M}$-stacking. All energies are given in relation to the lowest lying A exciton.}
\end{figure}

In figure \ref{fig:ordering}, the lowest lying exciton state for all layer configurations at the $R_h^M$-stacking is displayed. It is evident  that the $\Gamma$-K exciton is clearly the energetically lowest state. It is located about 300~meV below the lowest bright A exciton (intralayer K-K in MoS$_2$). Its purple color indicates that the $\Gamma$-K exciton is strongly hybridized reflecting a strong hole tunneling. The lowest lying K-K state is an interlayer exciton (red color). 
\begin{figure}[t!]
\hspace*{-1cm}  
\includegraphics[width=7.5cm]{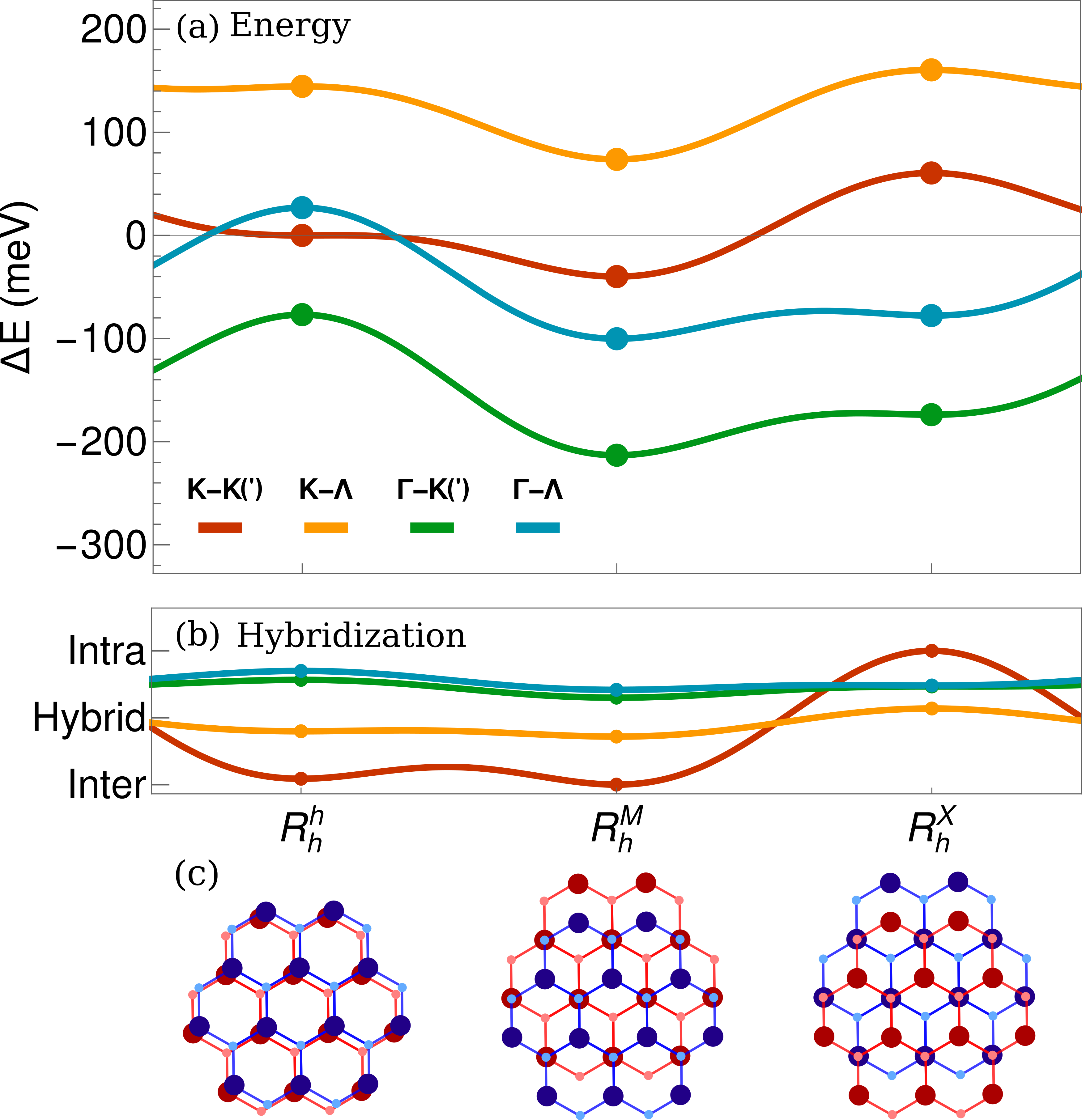}
\caption{\label{fig:bandstruct} $\bm{(a)}$ Lowest lying exciton state for each exciton valley configuration as a function of stacking for the MoS$_2$-WS$_2$ heterostructures on a SiO$_2$ substrate. All energies are expressed in relation to the bright K-K exciton at $R^h_h$. Note that K-K and K-K$^{\prime}$ as well as $\Gamma$-K and $\Gamma$-K$^{\prime}$ are almost degenerate. $\bm{(b)}$ Degree of hybridization for each exciton valley as a function of stacking. $\bm{(c)}$ Schematic for R-type stacking configurations, where blue indicates the WS$_2$ and red the MoS$_2$ layer.}
\end{figure}

Figure \ref{fig:bandstruct}a illustrates the lowest lying exciton state for each exciton valley as a function of stacking, while \autoref{fig:bandstruct}b shows the corresponding degree of hybridization. Energies and coefficients were computed for the three high-symmetry stackings $R_h^h, R^M_h, R^X_h$ (circles in \autoref{fig:bandstruct}a,b), whereas intermediate displacements have been interpolated with the fit function given in Appendix A. We predict that the $\Gamma$-K($^{\prime}$) excitons are the lowest lying states regardless of stacking. Note that the $\Gamma$-K$^{\prime}$ exciton is about 1 to 5~meV lower than the $\Gamma$-K state, this is mainly due to the slightly increased effective mass at the K$^{\prime}$ point compared to that of the K point. Furthermore, we find that the lowest K-K transition is given by interlayer excitons, except for the $R_h^X$-stacking, where the intralayer exciton gives the largest contribution to the lowest lying K-K exciton. The K-$\Lambda$ exciton, although very strongly hybridized, is located about 100~meV above the K-K exciton for all stackings. This is mostly due to the large energy difference between the $\Lambda$ and the K point band edges in MoS$_2$. The $\Gamma$-$\Lambda$ exciton interestingly exchanges place with the K-K exciton, at $R_h^h$ compared to $R_h^{M/X}$. This can  be ascribed to the change in interlayer distance, which strongly affects the hybridization and has been implicitly taken into account in the energies extracted from our DFT calculations as in the latter we allowed for relaxation of the interlayer spacing.

\begin{figure}[t!]
\hspace*{-1cm}  
\includegraphics[width=7.5cm]{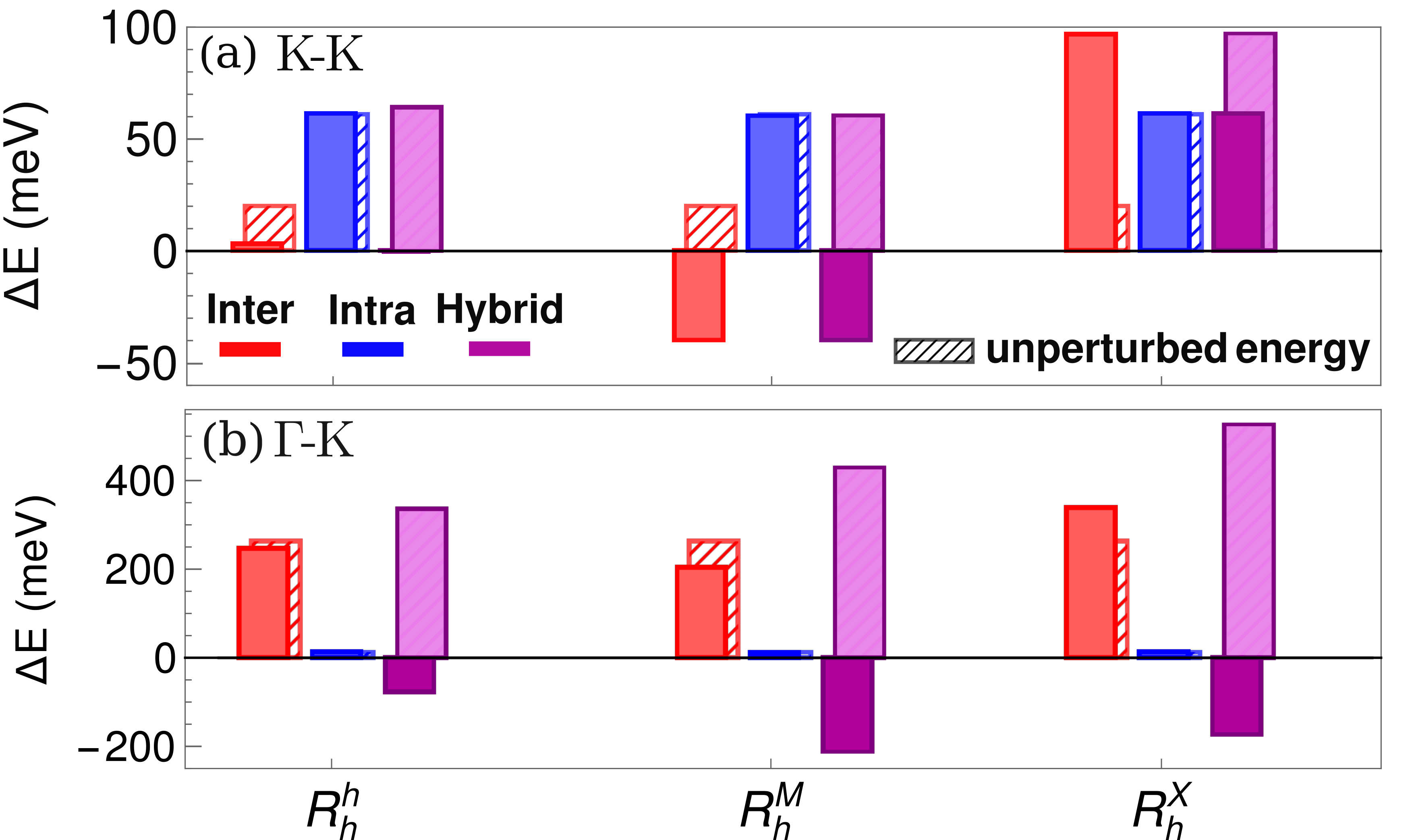}
\caption{\label{fig:microscopicCont} Microscopic contribution of the polarization-induced alignment shifts (red/blue bar) in relation to the unperturbed exciton energy (red/blue hatched bar) for the lowest lying intralayer (blue) and interlayer (red) $\bm{(a)}$ K-K and $\bm{(b)}$ $\Gamma$-K exciton in MoS$_2$-WS$_2$. The two final hybrid exciton states are indicated with light (repulsive hybrid) and dark purple bars (attractive hybrid). All energies are given in relation to the lowest lying hybrid K-K exciton in $R_h^h$-stacking.}
\end{figure}

By turning on and off certain parts of the Hamiltonian in \autoref{eq:ExcitonHamiltonian}) and consequently modifying the eigenvalue problem in \autoref{eq:EigenVal}, we can now disentangle the relative microscopic effects of hybridization and alignment shifts. Figure (\ref{fig:microscopicCont}) resolves these contributions for K-K and $\Gamma$-K excitons at various R-type stackings in MoS$_2$-WS$_2$. Here, the red (blue) hatched bars indicate the lowest lying unperturbed interlayer (intralayer) exciton and the filled red bar, the lowest lying interlayer (intralayer) exciton after applying the alignment shifts. The two purple bars display the energetic position for the two final hybrid exciton states including the effect of hybridization. 

\begin{table*}[t!]
\begin{tabular}{l*{6}{cld}}
\hline\hline
& \multicolumn{3}{l}{$\text{MoS}_2-\text{WS}_2$}
& \multicolumn{3}{l}{$\text{MoSe}_2-\text{WSe}_2$}
& \multicolumn{3}{l}{$\text{MoS}_2-\text{MoS}_2$}
& \multicolumn{3}{l}{$\text{WS}_2-\text{WS}_2$}
& \multicolumn{3}{l}{$\text{MoSe}_2-\text{MoSe}_2$}
& \multicolumn{3}{l}{$\text{WSe}_2-\text{WSe}_2$}
\\
\hline

$R_h^h$
&& $\Gamma$-K($^\prime$) & -140
&& K-K(I)                & -75
&& $\Gamma$-K($^\prime$) & -134
&& K-$\Lambda$           & -133
&& K-K                   & 0
&& K-$\Lambda$           & -158
\\
$R_h^M$
&& $\Gamma$-K($^\prime$) & \mathbf{-273}
&& K-K(I)                & \mathbf{-94}
&& $\Gamma$-K($^\prime$) & -304
&& $\Gamma$-$\Lambda$    & -241
&& K-$\Lambda$           & -6
&& K-$\Lambda$           & -209
\\

$R_h^X$
&& $\Gamma$-K($^\prime$) & -235
&& K-K                   & 0
&& $\Gamma$-K($^\prime$) & \mathbf{-304}
&& $\Gamma$-$\Lambda$    & -241
&& K-$\Lambda$           & -6
&& K-$\Lambda$           & -209
\\[3pt]
 
$H_h^h$
&& $\Gamma$-K($^\prime$) & -240
&& K-$\Lambda$           & -75
&& $\Gamma$-K($^\prime$) & -248
&& $\Gamma$-$\Lambda$    & \mathbf{-274}
&& K-$\Lambda$           & -12
&& K-$\Lambda$           & \mathbf{-216}
\\

$H_h^X$
&& $\Gamma$-K($^\prime$) & -218
&& K-K$^\prime$          & -45
&& $\Gamma$-K($^\prime$) & -251
&& $\Gamma$-$\Lambda$    & -237
&& K-K                   & 0
&& K-$\Lambda$           & -171
\\

$H_h^M$
&& $\Gamma$-K($^\prime$) & -162
&& K-K$^\prime$          & -51
&& $\Gamma$-K($^\prime$) & -160
&& $\Gamma$-$\Lambda$    & -109
&& K-K                   & 0
&& K-$\Lambda$           & -142
\\

\hline\hline
\end{tabular}
\caption{\label{tab:table1} The lowest lying exciton state for different TMD homo- and heterobilayers in different high-symmetry stacking configurations. The energy corresponds to the spectral distance (in meV) to the lowest lying A exciton (intralayer), in the corresponding stacking and material. For excitons denoted with K($^{\prime}$), the K$^{\prime}$-related exciton lies about 1 to 5~meV lower. The notation (I) indicates that it is an interlayer exciton.}
\end{table*}

We first focus on the microscopic contributions of hybridization and alignment shifts for the K-K exciton (\autoref{fig:microscopicCont}a). We find that the interlayer exciton is the energetically lowest state at $R_h^h$ and $R_h^M$ stacking (red bar). This is due to the large band offset between the layers (\autoref{fig:Schematic}a) and the alignment red-shift. Rather unexpectedly at $R_h^X$, the intralayer exciton becomes the energetically lowest state (blue bar). Here, the metal atoms of the upper layer (WS$_2$) are on top of the chalcogen atoms of the bottom layer (MoS$_2$) (\autoref{fig:bandstruct}c). This variation in atomic configuration leads to a change in direction of the layer polarization field, thus changing the nature of the energy shift relative to $R_h^h$. As a result, the lowest lying interlayer exciton becomes blue-shifted --- opposite to the red-shift for the $R_h^M$ stacking (solid vs hatched red bar), where the chalcogen atoms of the upper layer (WS$_2$) are on top of metal atoms of the bottom layer (MoS$_2$) (\autoref{fig:bandstruct}c). This blue-shift is large enough to compensate for the otherwise large band offset between the layers, thus making the intralayer exciton the lowest lying state. However, this effect is difficult to observe in PL experiments due to the small oscillator strength of the interlayer exciton and the overall dominant nature of $\Gamma$-K excitons in MoS$_2$-WS$_2$ (\autoref{fig:bandstruct}a). In addition, we observe that the K-K exciton is only very weakly affected by hybridization at $R_h^h$ and not at all at $R_h^{M/X}$ stacking (compare the purple bars with red and blue bar, respectively). The vanishing tunneling at $R_h^{M/X}$ is enforced by the orbital-symmetry of the states at the K point (\autoref{eq:TmatrixElement}). As a result, the variation in the energy for the K-K exciton is only due to the alignment shift. 

Considering the situation for the $\Gamma$-K exciton in \autoref{fig:microscopicCont}b, we immediately observe the significant role of hybridization that is reflected in the large red-shift of the dark purple bars. This is expected due to the strong tunneling of holes at the $\Gamma$ point. The strong variation in energy between different stackings for the hybrid $\Gamma$-K exciton mainly results from the varying interlayer distance, which has a strong impact on the tunneling strength (see \autoref{eq:Tunnelling}). However, also the stacking-dependent layer polarization field changes the energetic distance of the initial intra- and interlayer states and therefore indirectly influences the hybridization process.

We performed calculations for multiple TMD homo- and heterobilayers mapping out their entire exciton energy landscape. The results are summarized in \autoref{tab:table1} including the lowest lying exciton for each material and stacking and its energetic distance to the bright K-K exciton. In the MoS$_2$-WS$_2$ heterostructure, we predict the momentum-dark $\Gamma$-K($^\prime$) exciton to be the lowest regardless of stacking. The situation is completely different in MoSe$_2$-WSe$_2$, where the bright K-K exciton is the lowest in R-type stacking, the K-$\Lambda$ exciton in $H_h^h$ and the K-K$^{\prime}$ exciton in $H_h^{M/X}$-stacking. Interestingly, the K-$\Lambda$ exciton becomes the lowest due to the increased tunneling strength with the varying interlayer distance, while the K-K$^{\prime}$ exciton is the lowest due to the nature of the H-type stacking, where the spin ordering changes in one of the layers. This means that the K-K$^{\prime}$ exciton in H-type structures corresponds to K-K excitons in R-type stackings \cite{ruiz2019interlayer}. For TMD homobilayers, $R_h^M$ and $R_h^X$ are equivalent stackings just mirrored in the out-of-plane direction. Similarly to MoS$_2$-WS$_2$, the $\Gamma$-K($^\prime$) exciton is also the lowest state in MoS$_2$-MoS$_2$ bilayers due to the strong tunneling of holes around the $\Gamma$ point. In contrast, in WSe$_2$-WSe$_2$ the K-$\Lambda$ exciton is the lowest reflecting the efficient tunneling of electrons at the $\Lambda$ point, whereas the $\Gamma$-valley cannot compete due to the large energy separation to the K point in the monolayer case. Interestingly, we find a different lowest exciton in WS$_2$-WS$_2$ depending on the stacking: K-$\Lambda$ exciton at the $R_h^h$-stacking and $\Gamma$-$\Lambda$ exciton at $R_h^M$ and $R_h^X$, where the interlayer distance is reduced resulting in a larger red-shift of the $\Gamma$-$\Lambda$ exciton due to the hybridization.

\subsection{Photoluminescence}
So far we have been focusing on exciton energies in TMD homo- and heterobilayers revealing a rich landscape of various direct and indirect excitons. Now we investigate the optical response of these excitons by calculating PL spectra. In particular, we take into account direct and indirect recombination of excitons, the latter driven by emission and absorption of optical and acoustical phonons. To this end, we exploit the generalized Elliot formula for phonon-assisted PL derived in Refs.~\cite{brem2020phonon,D0NR02160A} reading
\begin{equation}\label{eq:PL}
\begin{split}
    I_{\sigma}(\omega)\propto \sum_{\eta \xi}\frac{|\widetilde{M}_{\sigma}^{\xi \eta}|^2}{(\mathcal{E}^{\xi}_{\eta 0}(S)-\hbar\omega)^2+(\gamma^{\xi}_{\eta}+\Gamma^{\xi^{\prime}}_{\eta})^2}\Big(\gamma^{\xi}_{\eta}N^{\xi}_{\eta 0}+\\\sum_{\substack{\xi^{\prime} \eta^{\prime}\\ \bm{q} j \pm}}|\widetilde{D}_{\xi \eta j \bm{0}}^{\xi^{\prime} \eta^{\prime} \bm{q}}|^2N^{\xi^{\prime}}_{\eta^{\prime}\bm{q}}n_{\bm{q}j}^{\pm}L(\mathcal{E}^{\xi^{\prime}}_{\eta^{\prime} \bm{q}}(S)\pm \Omega_{\bm{q} j}-\hbar\omega,\Gamma^{\xi^{\prime}}_{\eta^{\prime}})\Big),
\end{split}
\end{equation}
where $\eta(\eta^{\prime})$ is the hybrid exciton quantum number, $\sigma$ the polarization of the photon, $\xi(\xi^{\prime})$ the valley index, $\bm{q}$ the involved phonon momentum, $j$ the phonon mode and $\pm$ denotes phonon absorption ($+$) and emission ($-$). The first part of the equation describes the direct recombination of bright excitons within the light cone at the K point. Here, $L$ is a Cauchy–Lorentz distribution and $\widetilde{M}_{\sigma}^{\xi \eta}$ is the exciton-photon matrix element determining the oscillator strength of the hybrid excitons. Furthermore, $\mathcal{E}^{\xi}_{\eta \bm{Q}}(S)$ is the hybrid exciton energy as calculated by the eigenvalue problem given in \autoref{eq:EigenVal}. The radiative and non-radiative broadening are described by $\gamma^{\xi}_{\eta}$ and $\Gamma^{\xi^{\prime}}_{\eta}$, respectively. Since this work mainly focuses on the spectral position of the peaks, we account for these phenomenologically (Appendix D).

After optical excitation, excitons distribute throughout different valleys according to the Boltzmann distribution $N^{\xi}_{\eta\bm{q}}$. By emitting or absorbing a phonon,  momentum-dark excitons can become optically visible via (indirect) phonon-assisted recombination or in pump-probe spectra \cite{Berghauser2018PRB}. This higher-order process is determined by the second part of \autoref{eq:PL}, where $\widetilde{D}_{\xi \eta j \bm{0}}^{\xi^{\prime} \eta^{\prime} \bm{q}}$ is the exciton-phonon matrix element that takes into account the electron-phonon coupling deformation potentials (Appendix D) \cite{jin2014intrinsic,li2013intrinsic}. Furthermore, $\Omega_{\bm{q} j}$ denotes the phonon energy and $\tilde{n}_{\bm{q}j}^{\pm}=1/2\mp1/2+n_B(\Omega_{\bm{q} j})$ the phonon occupation, which is given by the Bose-Einstein distribution $n_B(\Omega_{\bm{q} j})$. In this equation we have neglected the unlikely higher-order process involving two phonons, which would be needed to scatter the $\Gamma$-$\Lambda$ exciton to the bright K-K state. 

\begin{figure}[t!]
\includegraphics[width=8.cm]{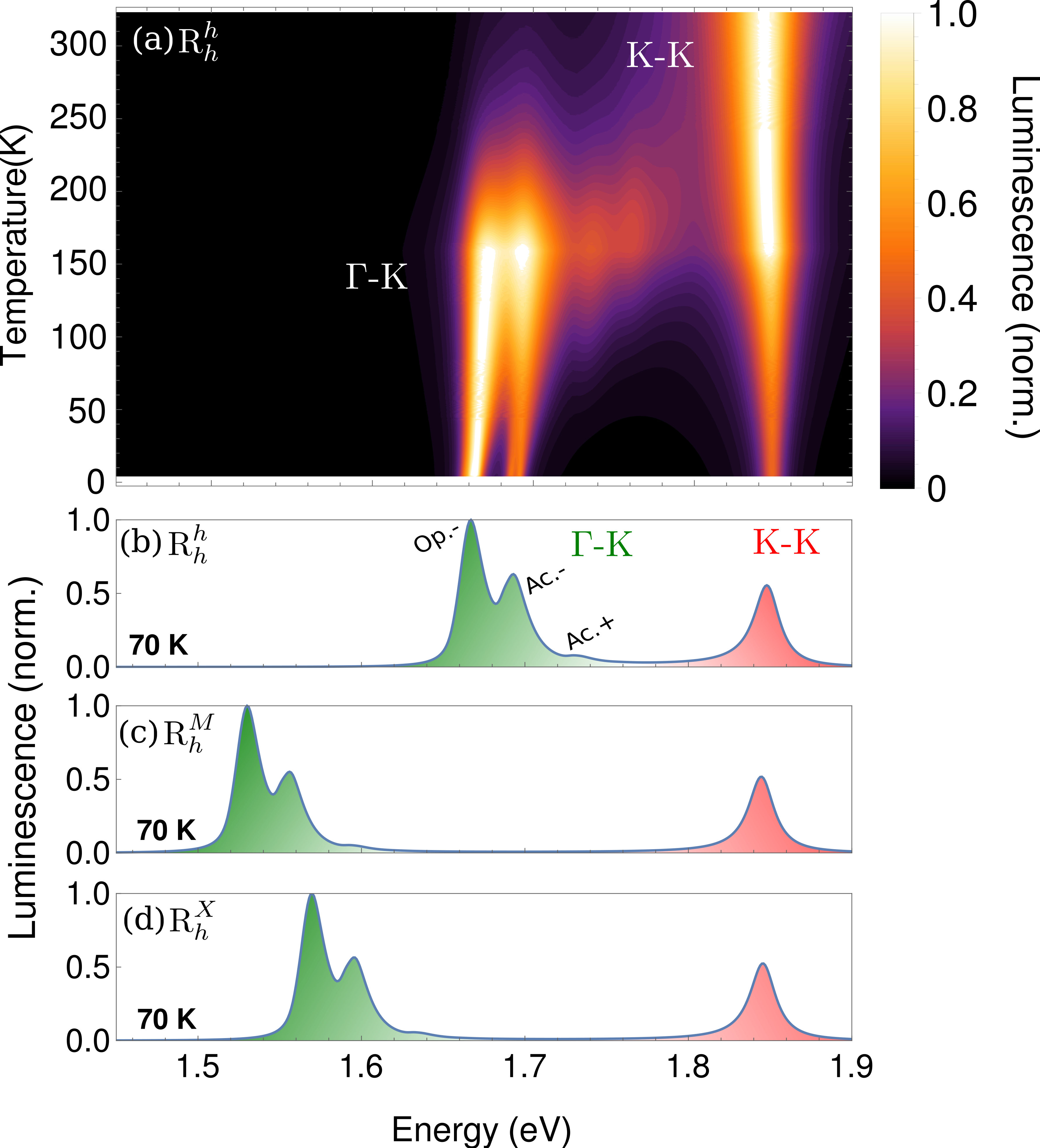}
\caption{\label{fig:PL}$\bm{(a)}$ Normalized photoluminescence spectra as a function of temperature and energy for $R_h^h$-stacking in the MoS$_2$-WS$_2$ heterostructure. PL spectra at 70 K for $\bm{(b)}$ $R_h^h$, $\bm{(c)}$ $R^M_h$ and $\bm{(d)}$ $R^X_h$ stacking. The green and the red shaded areas indicate the phonon sidebands of the $\Gamma$-K exciton and the K-K intralayer exciton, respectively. Phonon sidebands stemming from emission ($-$) and absorption ($+$) of optical (Op) and acoustical (Ac) phonons are labeled accordingly. }
\end{figure}

By evaluating \autoref{eq:PL} numerically for different TMD bilayers at various temperatures, we reveal the optical footprint of the exciton landscape for different materials at different stackings. Figure \ref{fig:PL}a shows the temperature-dependent PL in the exemplary case of MoS$_2$-WS$_2$ in $R_h^h$ stacking. Here the position of the A exciton is fixed to experimental values \cite{enalim2019restoring}. PL spectra from other TMD homo- and heterobilayers for different stackings are shown in the appendices. We find that at higher temperatures, the bright K-K exciton dominates the PL, while at lower temperatures, phonon sidebands from the $\Gamma$-K exciton become pronounced. They stem from emission of acoustical and optical phonons and are clearly separated at very low temperatures, while they overlap at intermediate temperatures reflecting the increased non-radiative broadening. In the temperature range around 150~K, we find the emergence of additional peaks stemming from phonon absorption. We can also see a clear asymmetrical broadening of the peaks at higher temperatures. This is due to the Boltzmann nature of the exciton distribution and is characteristic for the appearance of phonon sidebands.

To further understand the nature of the multi-peaked PL, we show the PL at 70~K at different stacking in \autoref{fig:PL}b-d. We identify the peak with the highest intensity as a phonon sideband stemming from the emission of an optical phonon. The next highest peak can be traced back to the emission of an acoustical phonon. These peaks have about equal contribution from their respective transverse and longitudinal components. There is a slight variation in the relative intensity of these peaks with stacking. This is due to the weak change in the degree of hybridization (\autoref{fig:microscopicCont}b), which means that the phonon contributions from each layer will be weighted differently. A small peak stemming from the acoustical absorption phonon mode is also visible at higher energies. The PL at the three different stackings is qualitatively the same. We observe a clear stacking-dependent red-shift of the phonon sidebands reflecting different hybridization of the $\Gamma$-K exciton (\autoref{tab:table1}).

Our results are in good agreement with experiments, where two predicted phonon sidebands have been observed about 300~meV below the bright K-K exciton in MoS$_2$-WS$_2$ \cite{enalim2019restoring}. This corresponds well to the calculated PL at $R_h^M$- and $R_h^X$-stacking, which is the energetically most favorable R-type stackings. The relative intensity of the two predicted phonon sidebands favoring the lower one agrees well with the experimental observation. An indirect dark exciton has also been observed in MoS$_2$-MoS$_2$ \cite{cao2019phonon,zhao2013origin} and WSe$_2$-WSe$_2$ \cite{jones2014spin,zhao2013origin} located about $300$ meV and $200$ meV below the bright exciton, respectively - in a very good agreement with our results summarized in \autoref{tab:table1}. Similar observations have been made also for WS$_2$-WS$_2$ including a peak roughly $200$ meV below the K-K exciton \cite{zheng2015coupling}. This is in good agreement with the predicted phonon sideband from the K-$\Lambda$ exciton in our calculations. Overall, our work provides microscopic insight into the exciton landscape and the optical signatures for a variety of TMD bilayers. In particular, the energy landscapes obtained for different lateral displacements play an important role for the understanding of the spatially dependent stacking domains and resulting moir\'e potentials \cite{enaldiev2020stacking, lu2019modulated} in twisted van der Waals bilayers.  

\section{Conclusion}
We have presented a material specific and fully microscopic model revealing the exciton landscape in TMD homo- and heterobilayers at different high-symmetry stackings. We identified the energetically lowest lying exciton state for each respective material and stacking. By combining the exciton density matrix formalism and density functional theory, we have gained microscopic insights into different contributions determining the hybrid exciton states including tunneling-induced hybridization and layer polarization-induced alignment shifts. Finally, we predict the optical footprint of the calculated exciton landscape in photoluminescence spectra including direct and phonon-assisted indirect exciton recombination processes. This allows us to uncover the exciton type behind the most prominent resonances for different TMD bilayers at different stackings. Furthermore, revealing the energy landscape at different stackings gives us insight into the expected behavior of the moir\'e potential at high-symmetry points in a twisted structure. Presenting the exciton landscape and its impact on optical spectra, our work can guide future theoretical and experimental studies in the growing field of van der Waals heterostructures. 

\section*{Acknowledgments}
This project has received funding from  Deutsche Forschungsgemeinschaft via CRC 1083 (project B09), the European Unions Horizon 2020 research and innovation programme under grant agreement no. 881603 (Graphene Flagship) as well as the Swedish Research Council (2018-06482, 2020-04935). Furthermore, we acknowledge the support from the Knut and Alice Wallenberg Foundation via the grant KAW 2019.0140 and Vinnova via the competence centre ``2D-TECH".
The computations were enabled by resources provided by the Swedish National Infrastructure for Computing (SNIC) at PDC and NSC partially funded by the Swedish Research Council through grant agreement no. 2018-05973.

\section*{Appendix A: Layer polarization-induced alignment shift}
The matrix elements governing the layer polarization-induced alignment shift of the single particle states are given by
\begin{equation}
\Delta\varepsilon^{\alpha\mathbf{k}}_l(S)= \int d\mathbf{r} n_{l,\alpha\mathbf{k}}(\mathbf{r}) \delta V_\text{pol}(\mathbf{r}).
\end{equation}
They were computed within the PAW formalism, where $n_{l,\alpha\mathbf{k}}$ is the orbital density of the state $|\alpha\mathbf{k}\rangle$ in the monolayer $l$. Here, $\alpha=(\lambda,\xi)$ is a compound index containing $\lambda=(c,v)$ as the band index and $\bm{\xi}$ as the valley index. To this end, $\delta V_\text{pol}$ is the solution to the Poisson equation for the electron density difference $\delta n$
$(\nabla^2 \delta V_\text{pol} -\delta n = 0)$ with $\delta n = n_{1,2} - n_1 - n_2$, with the subscript indicating the layer. Band renormalizations and monolayer energies, i.e $\tilde{E}^{\alpha}_{l\bm{k}}(S)=E^{\alpha }_{l\bm{k}}+\Delta\varepsilon^\lambda_l(S)$ (see \autoref{eq:Hamiltonian}), are extracted from DFT calculations for various homo- and heterobilayers and  are summarized in \autoref{tab:table1A} for R-type structures and in \autoref{tab:table2} for H-type structures. The indices $\mathbf{k}$ and $\bm{\xi}$ have been dropped from the alignment shift $\Delta\varepsilon^\lambda_l(S)$, since the latter is a valley-independent energy renormalization of the entire band structure. The intermediate displacements have been interpolated with a continuous fit function given by \cite{brem2020tunable},
\begin{equation}\label{eq:fitting}
    \tilde{E}^{\lambda}_{l}(S)=\mathcal{E}^{\lambda}_{l}+\Big(\alpha^{\lambda}_l+\beta^{\lambda}_l e^{\frac{2\pi i\sigma_{1-l}}{3}}\Big)\sum_{n=0}^{2}e^{i\text{C}^{(n)}_3\bm{G}_l^{(0)}\cdot \bm{D}_l(S)},
\end{equation}

\begin{figure*}[t!]
\hspace*{-1cm}  
\includegraphics[width=\linewidth]{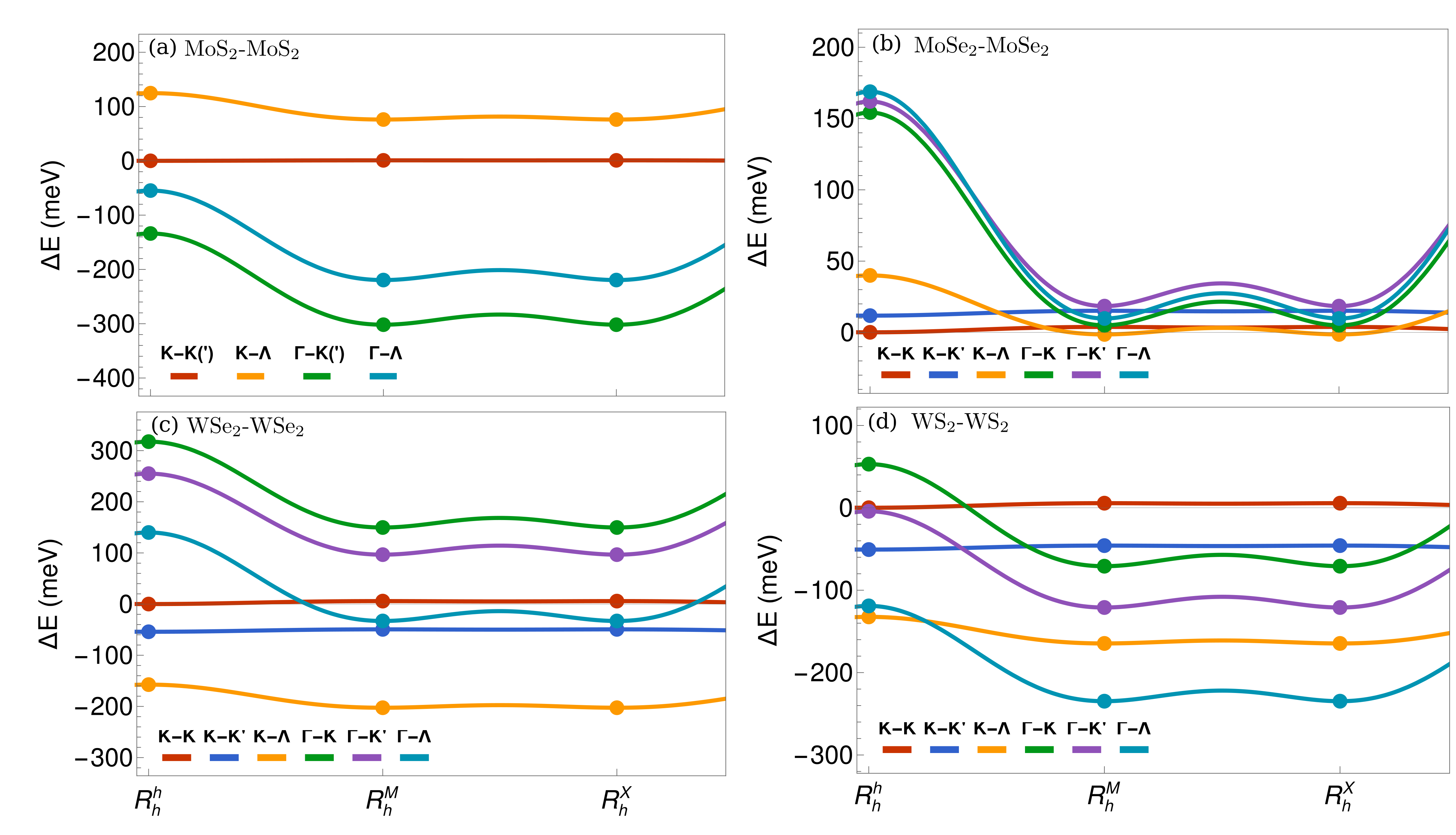}
\caption{\label{fig:bandstruct2} Lowest lying exciton state for each valley configuration as a function of stacking for all four TMD homobilayers: $\bm{(a)}$ MoS$_2$-MoS$_2$, $\bm{(b)}$ MoSe$_2$-MoSe$_2$, $\bm{(c)}$ WSe$_2$-WSe$_2$ and $\bm{(d)}$ WS$_2$-WS$_2$. All energies are expressed in relation to the K-K exciton at $R^h_h$ stacking for the respective material.  }
\end{figure*}

where $\mathcal{E}^{\lambda}_{l}$ is the stacking-independent monolayer energy and $\sigma_l=(-1)^{l}$ for H-type stacking, otherwise $\sigma_l=1$. Furthermore, $\alpha^{\lambda}_l$ and $\beta^{\lambda}_{l}$ determine the stacking-dependent shift. These parameters are fitted to the band structure for the K point, obtained from first-principle calculations $\tilde{E}^{\lambda \text{K}}_{l\bm{0}}(S)=E^{\alpha }_{l\bm{0}}+\Delta\varepsilon^\lambda_l(S)$, where the momentum $\bm{k}=\bm{0}$, i.e the minimum of the parabolic dispersion. For the correct energetic position of other valleys, one has to take into account the spectral valley separation and spin orbit splitting Ref. \cite{Korm_nyos_2015} listed in \autoref{tab:table_kormanyos}. The fitted parameters for all materials can be found in Tables \ref{tab:table3} and  \ref{tab:table4}. The tunneling strength as expressed in \autoref{eq:Tunnelling} in the main part of the manuscript is given in \autoref{tab:table5} for different homo- and heterobilayers.

\section*{Appendix B: Exciton Hamiltonian}
We transform the Hamilton operator given in \autoref{eq:Hamiltonian} into an exciton basis \cite{D0NR02160A,brem2020tunable}. For this purpose, we expand the electronic creation and annihilation operators in pair operators $P^{\dagger}_{ik,j\bm{k}^{\prime}}=a^{c \dagger}_{i \bm{k}}a^{v}_{j \bm{k}^{\prime}}$ in the low density limit (with $i=(\xi_i,l_i)$ as a compound index) yielding 
\begin{equation}
\begin{split}
a^{c \dagger}_{i \bm{k}}a^{c}_{j \bm{k}^{\prime}}\approx\sum_{m\bm{p}}P^{\dagger}_{i\bm{k},m\bm{p}}P_{j\bm{k}^{\prime},m\bm{p}}\\
a^{v \dagger}_{i \bm{k}}a^{v}_{j \bm{k}^{\prime}}\approx\delta^{ij}_{\bm{k}\bm{k}^{\prime}}-\sum_{m\bm{p}}P^{\dagger}_{m\bm{p},j\bm{k}^{\prime}}P_{m\bm{p},i\bm{k}}.
 \end{split}
\end{equation}
This allows us to expand the Hamiltonian in center-of-mass coordinates, using the exciton operators \cite{katsch2018theory,D0NR02160A} yielding
\begin{equation}
P^{\dagger}_{i\bm{k},j\bm{k}^{\prime}}=\sum_{\mu}X^{\mu\dagger}_{ij,\bm{k}-\bm{k}^{\prime}}\Psi^{\mu}_{ij}(\alpha_{ij}\bm{k}^{\prime}+\beta_{ij}\bm{k}).
\end{equation}
Here, $\mu$ is the exciton quantum number, which we set to 1s, since we are focusing on the lowest lying exciton states. Furthermore, $\Psi^{\mu}_{ij}(\bm{k})$ is the exciton wavefunction as obtained from the solution of the Wannier equation \cite{ovesen2019interlayer}. Here, we have introduced $\alpha_{ij}=m^c_i/(m^c_i+m^v_j)$ and $\beta_{ij}=m^v_j/(m^c_i+m^v_j)$. This consequently results in an exciton Hamiltonian 
\begin{equation}\label{eq:ExcitonHamiltonian2}
\begin{split}
    H=\sum_{\substack{\xi\bm{Q}\\LL^{\prime}}}( E^{\xi S}_{L\bm{Q}}X^{\xi \dagger}_{L  \bm{Q}}X^{\xi }_{L  \bm{Q}}\delta_{LL^{\prime}}+T_{LL^{\prime}}^{\xi}(S)X^{\xi\dagger}_{L\bm{Q}}X^{\xi}_{L^{\prime}   \bm{Q}}),
        \end{split}
\end{equation}
where $\bm{Q}=\bm{k}-\bm{k}^{\prime}$ is the center-of-mass momentum, $\xi=(\xi_e,\xi_h)$ the exciton valley index  and $L(L)^{\prime}$ a compound index $L=(l_e,l_h)$  replacing the generic indices $i,j$. Here, we have assumed an untwisted structure and thus no momentum transfer is involved. The tunneling matrix element is given by
\begin{equation}
\begin{split}
T_{LL^{\prime}}^{\xi}(S)=\mathcal{F}^\xi_{LL^{\prime}}(T_{l_el_e^{\prime}}^{c \xi_e }(S)\delta_{l_e,l_e^{\prime}-1}\delta_{l_h,l_h^{\prime}}\\
-T_{l_hl_h^{\prime}}^{v \xi_h }(S)\delta_{l_h,l_h^{\prime}-1}\delta_{l_e,l_e^{\prime}}).    
 \end{split}
\end{equation}
The  matrix elements for the conduction and valance band $T_{ll^{\prime}}^{c(v) \xi }(S)$ are described in the main part of the manuscript. Finally, the exciton form factors $\mathcal{F}^\xi_{LL^{\prime}}$ are  given by
\begin{equation}
    \mathcal{F}^\xi_{LL^{\prime}}=\sum_{\bm{k}}\Psi_{\xi L}^*(\bm{k})\Psi_{\xi L^{\prime}}(\bm{k}).
\end{equation}
By expanding the exciton Hamiltonian into hybrid exciton operators, we can derive an eigenvalue problem, which diagonalizes the Hamiltonian and yields the final hybrid exciton energies as shown in the main text. 

\begin{figure}[b!]
\hspace*{-1cm}  
\includegraphics[width=\linewidth]{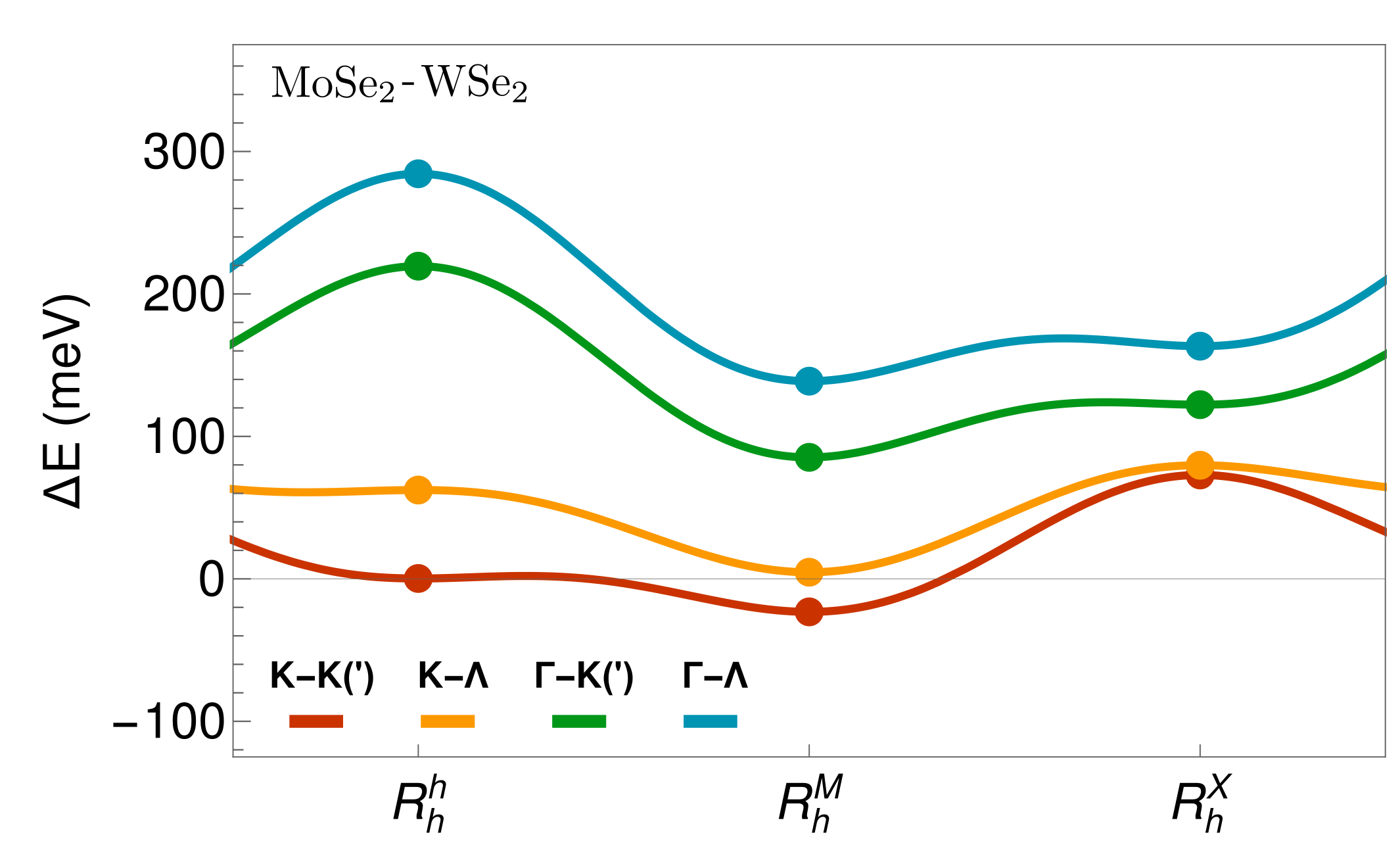}
\caption{\label{fig:bandstructhetero} Lowest lying exciton state for each valley configuration as a function of stacking for the MoSe$_2$-WSe$_2$ heterostructure.}
\end{figure}

\section*{Appendix C: Exciton energy landscape}
Here, we discuss the exciton energy landscape for the four homobilayers MoS$_2$, WS$_2$,WSe$_2$, MoSe$_2$ and the heterostructure MoSe$_2$-WSe$_2$ that have not been covered in the main text. Figure \ref{fig:bandstruct2}a shows the exciton landscape for  $\text{MoS}_2-\text{MoS}_2$ for the three R-type high-symmetry stackings. Similar to the heterostructure $\text{MoS}_2-\text{WS}_2$ discussed in the main text, in this homobilayer  $\Gamma$-K exciton is the energetically lowest state for all stackings. The $\Gamma$-$\Lambda$ exciton is also significantly red-shifted, but due to the large valley separation of the $\Lambda$ point, it still remains above the $\Gamma$-K exciton.

\begin{table*}
\begin{ruledtabular}
\begin{tabular}{c|cccc}
& $\text{WSe}_2$ & $\text{MoSe}_2$ & $\text{WS}_2$ & $\text{MoS}_2$\\
\hline
\hline
K$^{\prime}_v$ & -462 meV & -184 meV & -425 meV & -148 meV\\
K$^{\prime}_c$ & -37 meV & 22 meV & -31 meV & 3 meV\\
$\Lambda_c$ & -5 meV & 163 meV & 27 meV & 246 meV\\
$\Gamma_v$ & -506 meV & -329 meV & -269 meV & -46 meV\\
\hline
\end{tabular}
\end{ruledtabular}
\caption{\label{tab:table_kormanyos} Spectral valley separation and spin-orbit splitting at the K point as listed in Ref. \cite{Korm_nyos_2015}.}
\end{table*}

The exciton landscape for the $\text{MoSe}_2-\text{MoSe}_2$ homobilayer is illustrated in Figure \ref{fig:bandstruct2}b. Here, we can see that the K-K exciton is the lowest at $R_h^h$, while K-$\Lambda$ becomes the lowest at $R_h^M$ and $R_h^X$. This reflects the increased tunneling probability of electrons when the interlayer distance decreases at the latter two stacking configurations.

In the $\text{WSe}_2-\text{WSe}_2$ homobilayer shown in \autoref{fig:bandstruct2}c, we can instead see that the K-$\Lambda$ exciton is the lowest state. This is mostly due to the very strong tunneling of electrons around the $\Lambda$ point. In $\text{WSe}_2$, the $\Gamma$ point is very far away from the global valence band maximum and even though it has a larger tunneling strength  than at the $\Lambda$ point (see \autoref{tab:table2}), the associated $\Gamma$-K/$\Lambda$ excitons remain above K-$\Lambda$ states. Due to the significant spin-orbit splitting in the conduction band at the K point, we have also included the K$^{\prime}$-related excitons. 

\begin{figure}[b!]
\hspace*{-1cm}  
\includegraphics[width=\linewidth]{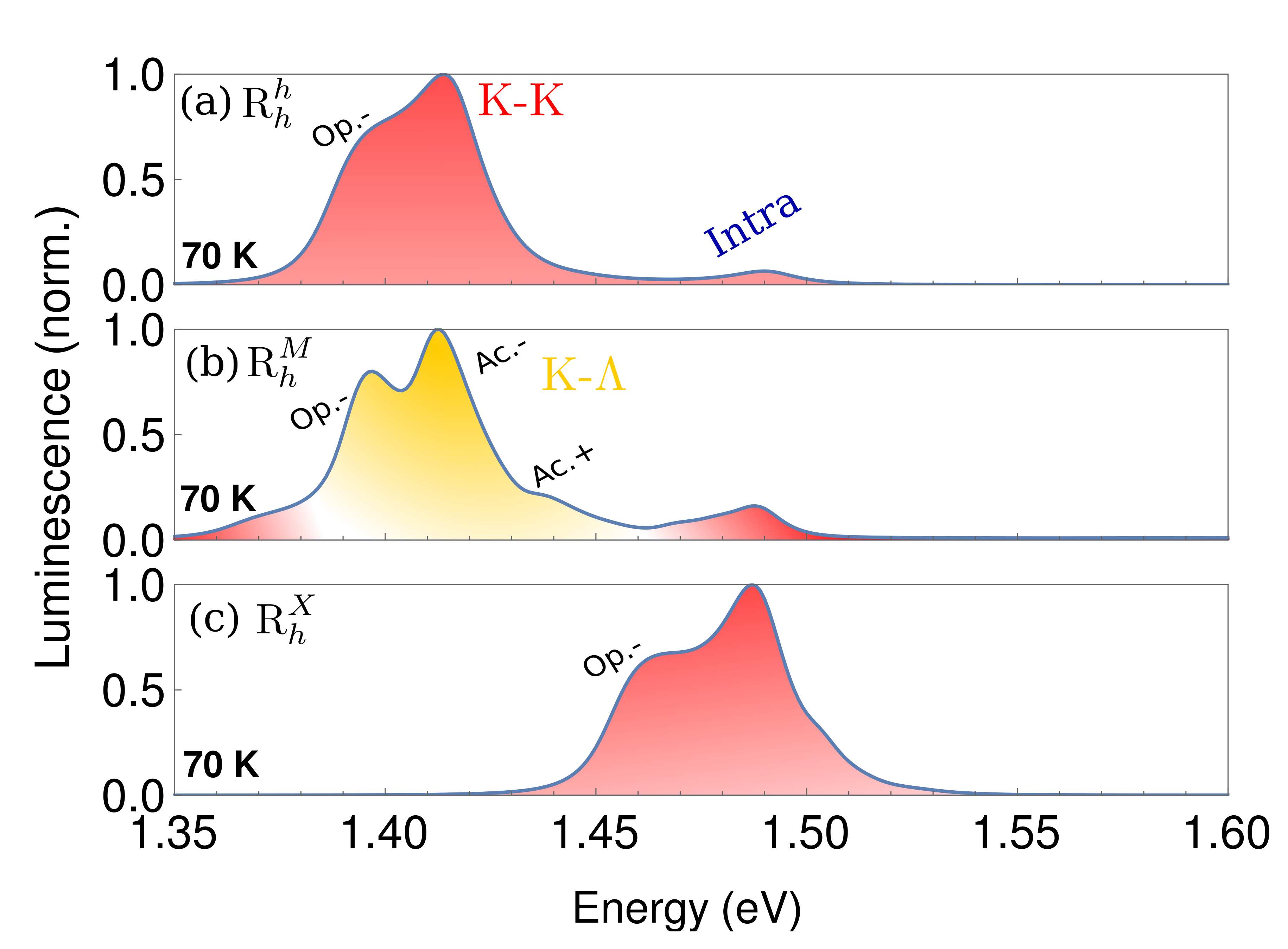}
\caption{\label{fig:PLhetero} Photoluminescence spectra for the  MoSe$_2$-WSe$_2$ heterostructure at 70 K.}
\end{figure}

Figure \ref{fig:bandstruct2}d illustrates the stacking-dependent exciton energies for the $\text{WS}_2-\text{WS}_2$ homobilayer. Here, we find that $\Gamma$-$\Lambda$ excitons are the energetically lowest  states. In this material, the  $\Gamma$ ($\Lambda$) point of the monolayer is energetically much closer to the valence (conduction) band edge at the K point. Consequently, considering the strong hybridization in these valles, the $\Gamma$-$\Lambda$ exciton becomes by far the lowest lying exciton in $R_h^{M/X}$-stacking. However, we observe  that at $R_h^h$ stacking, the K-$\Lambda$ is the lowest state instead. This is due to the increased interlayer distance at these stackings. 

For the four homobilayers discussed above, we can see no change between the $R_h^M$ and $R_h^X$-stacking, since these stackings are mirror-symmetric for a homobilayer. We also observe a very strong hybrid nature for the low lying dark excitons ($\Gamma$-K, $\Gamma$-$\Lambda$,  K-$\Lambda$) in comparison to the heterostructure discussed in the main part. This is mostly due to the lack of a band offset in homobilayers, which consequently reduces the detuning of the intra- and interlayer exciton and thus enhances the layer mixing. 

In figure \ref{fig:bandstructhetero} the exciton landscape for the heterostructure $\text{MoSe}_2-\text{WSe}_2$ is shown. In contrast to $\text{MoS}_2-\text{WS}_2$, which is discussed in the main part of the manuscript, the K-K exciton is the lowest lying exciton, regardless of stacking. This is mainly due to the large valley separation for $\Gamma$ (see \autoref{tab:table_kormanyos}) and the large band offset between the different materials. We can also see that the K-$\Lambda$ exciton lies very close to the bright K-K exciton, especially at $R_h^M$ and $R_h^X$ stackings.

\section*{Appendix D: Photoluminescence spectra}
As described in the main part of the manuscript, the photoluminescence (PL) spectra are calculated by considering both the resonant emission of photons from excitons and phonon-assisted recombination of excitons as a higher-order process. The exciton-photon matrix element which governs the direct photoemission reads \cite{brem2020phonon,D0NR02160A},
\begin{equation}
    \widetilde{M}_{\sigma}^{\xi \eta}=\sum_{L}M^{L \xi}_{\sigma}\mathcal{C}^{\xi \eta}_{L}(\bm{0})\delta_{\xi_e,\xi_h},
\end{equation}

where $M^{L \xi}_{\sigma}=\kappa^{\text{rad}}_L\sum_{\bm{k}} \Psi^{\xi}_L(\bm{k})$ is the oscillator strength and $\kappa^{\text{rad}}_L$ the square root of the radiative decay rate  \cite{Merkl2019}. Furthermore, $\delta_{\xi_e,\xi_h}$ ensures that only bright excitons are taken into account in the completely resonant case. 

\begin{figure*}[t!]
\hspace*{-1cm}  
\includegraphics[width=\linewidth]{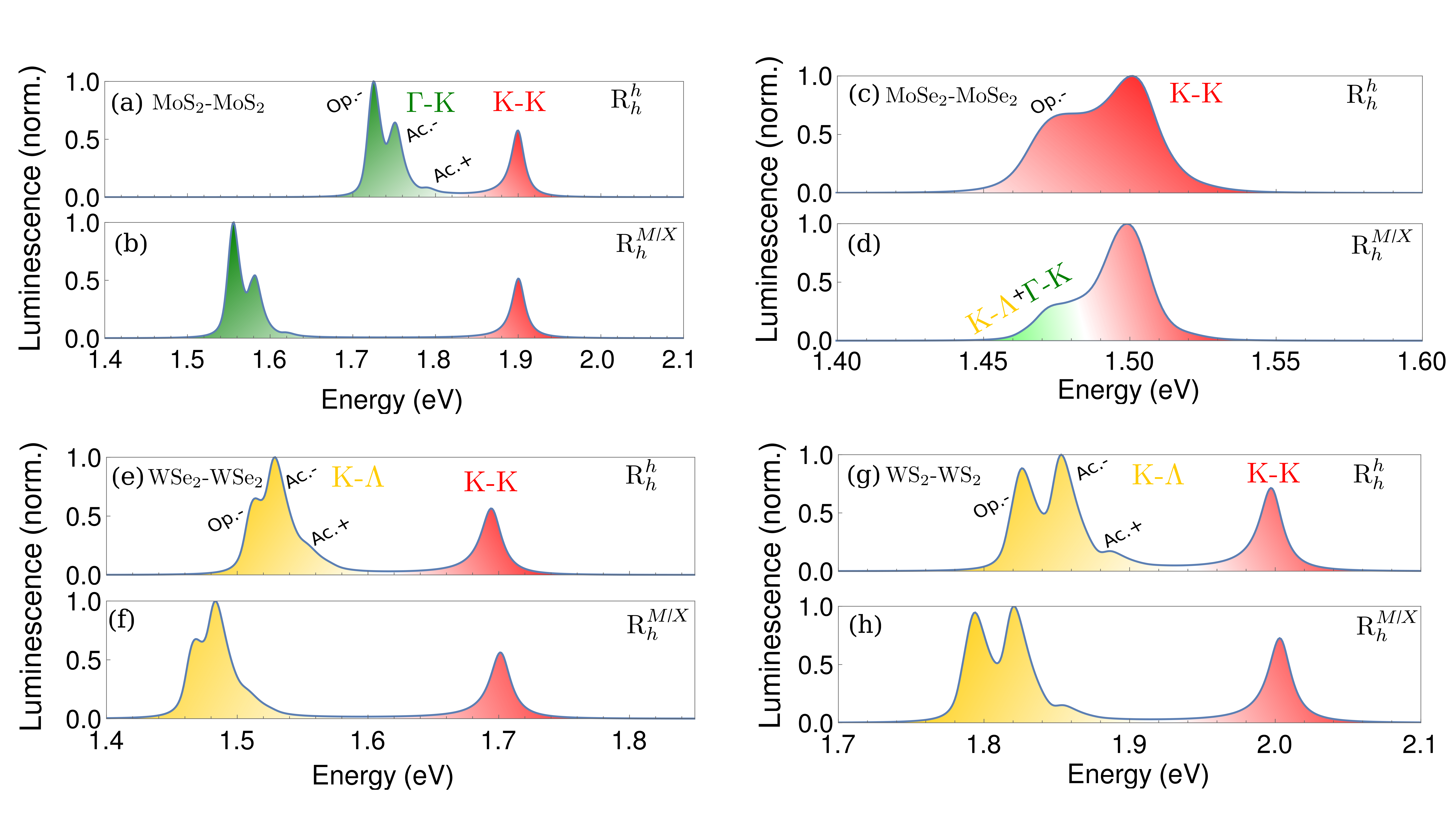}
\caption{\label{fig:PL2} Photoluminescence spectra for the four TMD homobilayers at 70 K at $R_h^h$ and $R_h^{M/X}$stacking: $\bm{(a)}$-$\bm{(b)}$ MoS$_2$-MoS$_2$, $\bm{(c)}$-$\bm{(d)}$ MoSe$_2$-MoSe$_2$, $\bm{(e)}$-$\bm{(f)}$ WSe$_2$-WSe$_2$ and $\bm{(g)}$-$\bm{(h)}$ WS$_2$-WS$_2$.}
\end{figure*}

For the indirect, phonon-assisted exciton recombination, the exciton-phonon matrix element $\widetilde{D}_{\xi \eta j 0}^{\xi^{\prime} \eta^{\prime} \bm{q}}$ plays the crucial role and is given by \cite{brem2020phonon,D0NR02160A}
\begin{equation}\label{eq:PhononExcitonMatrixElement}
\begin{split}
\widetilde{D}_{\xi \eta j \bm{0}}^{\xi^{\prime} \eta^{\prime} \bm{q}}=\sum_{LL^{\prime}}\Big(\mathcal{G}^{c j}_{\xi L,\xi^{\prime}L^{\prime}}(\bm{q}-\Delta\bm{\xi})\delta_{\xi_h\xi_h^{\prime}}\\
-\mathcal{G}^{v j}_{\xi L,\xi^{\prime}L^{\prime}}(\bm{q}-\Delta\bm{\xi})\delta_{\xi_e\xi_e^{\prime}}\Big)
\mathcal{C}^{\xi \eta}_{L}(\bm{0})\mathcal{C}^{\xi^{\prime} \eta^{\prime}}_{L}(\bm{q})\delta_{l_e,l^{\prime}_e}\delta_{l_h,l^{\prime}_h}.
 \end{split}
\end{equation}
Here, $\mathcal{C}^{\xi \eta}_{L}(\bm{\bm{q}})$ is the mixing coefficient and it holds $\mathcal{C}^{\xi \eta}_{L}(\bm{0+\bm{q}})\approx\mathcal{C}^{\xi \eta}_{L}(\bm{0})$ for untwisted structures, since the mixing coefficient only weakly vary due to similar masses of intra- and interlayer excitons. The transferred valley momentum $\Delta\bm{\xi}=(\xi^{\prime}_e-\xi^{\prime}_h-\xi_e+\xi_h)$ ensures the correct transformation between the globally defined phonon momentum and the exciton/electron momenta defined in valley local coordinates. The selection rule $\delta_{l,l^{\prime}}$ enforces that electron-phonon scattering takes place within one monolayer and that no tunneling occurs during this process. This assumption holds, as the interlayer atomic interaction is given by the van der Waals interaction, which is very weak in comparison to the in-plane atomic bonds. As a result,  it is highly unlikely for phonons to tunnel during this process. The exciton-phonon matrix elements $\mathcal{G}^{c(v)j}_{\xi L,\xi^{\prime}L^{\prime}}(\bm{q})$ for the conduction and valance band are given by  \cite{brem2020phonon,D0NR02160A},
\begin{equation}\label{eq:PhononExcitonMatrixElementBands}
\begin{split}
\mathcal{G}^{cj}_{\xi L,\xi^{\prime}L^{\prime}}(\bm{q})=g^{c j}_{\xi_e\xi_e^{\prime}}(\bm{q})\mathcal{F}^{\xi\xi'}_{LL^{\prime}}(\beta^{\xi\xi'}_{LL^{\prime}}\bm{q})\\
\mathcal{G}^{vj}_{\xi L,\xi^{\prime}L^{\prime}}(\bm{q})=g^{v j}_{\xi_h\xi_h^{\prime}}(\bm{q})\mathcal{F}^{\xi\xi'}_{LL^{\prime}}(-\alpha^{\xi\xi'}_{LL^{\prime}}\bm{q}).
\end{split}
\end{equation}
Here, $g^{c(v) j}_{\xi\xi^{\prime}}(\bm{q})$ are the electron-phonon coupling deformation potentials, calculated via ab-inito calculations in \cite{jin2014intrinsic,li2013intrinsic}. The intravalley scattering with acoustic phonons is approximated as linear in momentum, otherwise the coupling is given by a constant. Similarly the phonon dispersion in \autoref{eq:PL} in the main text is approximated as a constant (according to the Einstein model), while the long range acoustic phonons (intravalley scattering) are approximated as linear in momentum (Debye model). The values for the phonon dispersion is also calculated via ab-inito calculations in \cite{jin2014intrinsic,li2013intrinsic}.

The radiative $\gamma^{\xi}_{\eta}\approx1 \text{ meV}$ and non-radiative broadening $\Gamma^{\xi^{\prime}}_{\eta}$ of the exciton resonances are treated phenomenologically. The non-radiative broadening is approximated with a linear temperature dependence in accordance with Ref. \cite{selig2016excitonic,D0NR02160A} with $\Gamma^{\text{Bright}}_{\eta}(T)=5.0 \text{ meV}+0.05\cdot10^{-3}\text{ (meV/K) }T$ and $\Gamma^{\text{Dark}}_{\eta}=\Gamma^{\text{Bright}}_{\eta}/2.0$. The non-radiative broadening of dark excitons is significantly smaller, since they are lower in energy and therefore have less relaxation channels for scattering with phonons.

In \autoref{fig:PLhetero}, the Pl spectra for the heterostructure MoSe$_2$-WSe$_2$ is illustrated. Here we can see that the spectra is dominated by the bright K-K exciton at the $R^h_h$ and $R_h^X$ stacking. At $R^h_h$ we also observe a small peak from the A exciton. A special case arises when considering $R^M_h$-stacking: The lowest lying exciton is the interlayer K-K exciton, but due to its weak oscillator strength, the PL is dominated by the K-$\Lambda$ exciton instead. 

Figure \ref{fig:PL2}a-h show the PL spectra for the four homobilayers (MoS$_2$-MoS$_2$, WSe$_2$-WSe$_2$, WS$_2$-WS$_2$, MoSe$_2$-MoSe$_2$) for different stackings. The spectrally lowest $\Gamma$-K and K-$\Lambda$ resonances are marked green and yellow, respectively. We observe  that these momentum-dark excitons clearly dominate the PL at 70 K for MoS$_2$-MoS$_2$, WSe$_2$-WSe$_2$ and WS$_2$-WS$_2$. In the case of WS$_2$-WS$_2$, the actually lowest $\Gamma$-$\Lambda$ exciton is not visible, since here an unlikely two-phonon process is required for the indirect exciton recombination. In the case of MoSe$_2$-MoSe$_2$, the bright K-K exciton dominates with a low-energy shoulder stemming from the momentum dark $\Gamma$-K and K-$\Lambda$ excitons at $R^M_h$ and $R^X_h$ stackings.

\begin{table*}
\begin{ruledtabular}
\begin{tabular}{c|cccccc}
\multicolumn{2}{c}{} & $E^{\alpha }_{l\bm{0}}$ & \multicolumn{3}{c}{$\Delta\varepsilon^\lambda_l(S)$}  \\
\multicolumn{2}{c}{} &  & $R_h^h$ & $R_h^M$ & $R_h^X$ \\
\hline
\hline
\multirow{4}{*}{$\text{MoS}_2-\text{WS}_2$}& K-Mo(v) & -6157.1 meV & 7.7 meV & -14.1 meV & 52.9 meV \\
& K-W(v) & -5932.5 meV & -8.3 meV & 10.4 meV & -58.1 meV \\
& K-Mo(c) & -4371.3 meV & 6.8 meV & -15.7 meV & 52.8 meV \\
& K-W(c) & -3984.3 meV & -9.2 meV & 9.6 meV & -60.3 meV \\
\hline
\multirow{4}{*}{$\text{MoSe}_2-\text{WSe}_2$} & K-Mo(v) & -5555.1 meV & 10.5 meV & -6 meV & 46.5 meV \\
& K-W(v) & -5341.3 meV & -11.5 meV & 1.8 meV & -51.6 meV \\
& K-Mo(c) & -3994.8 meV & 9.5 meV & -8.1 meV & 46.7 meV \\
& K-W(c) & -3646.7 meV & -12.7 meV & 1.1 meV & -54.3 meV \\
\hline
\multirow{4}{*}{$\text{MoS}_2-\text{MoS}_2$} & K-L$_1$(v) & -6156.9 meV & -0.2 meV & -35.8 meV & 31.6 meV \\
& K-L$_2$(v) & -6156.9 meV & -0.2 meV & 31.6 meV & -35.8 meV \\
& K-L$_1$(c) & -4371.0 meV & -1.1 meV & -37.5 meV & 31.2 meV \\
& K-L$_2$(c) & -4371.0 meV & -1.1 meV & 31.2 meV & -37.5 meV \\
\hline
\multirow{4}{*}{$\text{MoSe}_2-\text{MoSe}_2$} & K-L$_1$(v) & -5555.2 meV & -0.4 meV & -29.5 meV & 25.4 meV \\
& K-L$_2$(v) & -5555.2 meV & -0.4 meV & 25.1 meV & -30.1 meV \\
& K-L$_1$(c) & -3994.8 meV & -1.4 meV & -31.7 meV & 25.4 meV \\
& K-L$_2$(c) & -3994.8 meV & -1.4 meV & 25.1 meV & -32.3 meV \\
\hline
\multirow{4}{*}{$\text{WS}_2-\text{WS}_2$} & K-L$_1$(v) & -5935.8 meV & -0.7 meV & -29.0 meV & 29.6 meV \\
& K-L$_2$(v) & -5935.8 meV & -0.7 meV & 26.4 meV & -33.4 meV \\
& K-L$_1$(c) & -3995.0 meV & -1.6 meV & -31.0 meV & 29.1 meV \\
& K-L$_2$(c) & -3995.0 meV & -1.6 meV & 25.7 meV & -35.4 meV \\
\hline
\multirow{4}{*}{$\text{WSe}_2-\text{WSe}_2$} & K-L$_1$(v) & -5342.9 meV & -0.5 meV & -30.2 meV & 25.0 meV \\
& K-L$_2$(v) & -5342.9 meV & -0.5 meV & 25.1 meV & -30.2 meV \\
& K-L$_1$(c) & -3651.9 meV & -1.6 meV & -32.6 meV & 24.7 meV \\
& K-L$_2$(c) & -3651.9 meV & -1.6 meV & 24.9 meV & -32.6 meV \\
\hline
\end{tabular}
\end{ruledtabular}
\caption{\label{tab:table1A} Stacking-dependent layer polarization-induced alignment shift of the electronic bandstructure and the monolayer energies for \it{R-type} structures as extracted from DFT calculations, i.e $\tilde{E}^{\alpha}_{l\bm{k}}(S)=E^{\alpha }_{l\bm{k}}+\Delta\varepsilon^\lambda_l(S)$.}
\end{table*}

\begin{table*}
\begin{ruledtabular}
\begin{tabular}{c|cccccc}
\multicolumn{2}{c}{} & $E^{\alpha }_{l\bm{0}}$ & \multicolumn{3}{c}{$\Delta\varepsilon^\lambda_l(S)$}  \\
\multicolumn{2}{c}{} &  & $H_h^h$ & $H_h^X$ & $H_h^M$ \\
\hline
\hline
\multirow{4}{*}{$\text{MoS}_2-\text{WS}_2$} & K-Mo(v) & -6156.9 meV & 18.6 meV & 15.4 meV & 8.8 meV \\
& K-W(v) & -5932.5 meV & -22.0 meV & -21.0 meV & -8.9 meV \\
& K-Mo(c) & -4371.1 meV & 19.1 meV & 14.3 meV & 7.5 meV \\
& K-W(c) & -3984.2 meV & -22.6 meV & -22.5 meV & -10.1 meV \\
\hline
\multirow{4}{*}{$\text{MoSe}_2-\text{WSe}_2$} & K-Mo(v) & -5554.4 meV & 20.0 meV & 17.5 meV & 12.0 meV \\
& K-W(v) & -5341.1 meV & -24.0 meV & -26.6 meV & -12.1 meV \\
& K-Mo(c) & -3993.3 meV & 20.6 meV & 16.7 meV & 10.4 meV \\
& K-W(c) & -3645.8 meV & -24.8 meV & -28.0 meV & -13.6 meV \\
\hline
\multirow{4}{*}{$\text{MoS}_2-\text{MoS}_2$} & K-L$_1$(v) & -6155.8 meV & -1.2 meV & -3.2 meV & -0.1 meV \\
& K-L$_2$(v) & -6155.8 meV & -1.2 meV & -3.2 meV & -0.1 meV \\
& K-L$_1$(c) & -4369.0 meV & -1.2 meV & -4.5 meV & -1.7 meV \\
& K-L$_2$(c) & -4369.0 meV & -1.2 meV & -4.5 meV & -1.7 meV \\
\hline
\multirow{4}{*}{$\text{MoSe}_2-\text{MoSe}_2$} & K-L$_1$(v) & -5554.4 meV & -2.1 meV &  -4.4 meV & 0.3 meV \\
& K-L$_2$(v) & -5554.4 meV & -2.1 meV & -4.4 meV & 0.3 meV \\
& K-L$_1$(c) & -3993.3 meV & -1.6 meV & -5.4 meV &  -1.3 meV \\
& K-L$_2$(c) & -3993.3 meV & -1.6 meV & -5.4 meV &  -1.3 meV \\
\hline
\multirow{4}{*}{$\text{WS}_2-\text{WS}_2$} & K-L$_1$(v) & -5934.5 meV & -1.3 meV & -3.1 meV & -0.2 meV \\
& K-L$_2$(v) & -5934.5 meV & -1.3 meV & -3.1 meV & -0.2 meV \\
& K-L$_1$(c) & -3993.6 meV & 0.0 meV & -4.4 meV & -1.3 meV \\
& K-L$_2$(c) & -3993.6 meV & 0.0 meV & -4.4 meV & -1.3 meV \\
\hline
\multirow{4}{*}{$\text{WSe}_2-\text{WSe}_2$} & K-L$_1$(v) & -5342.3 meV & -2.1 meV & -4.3 meV & -0.1 meV \\
& K-L$_2$(v) & -5342.3 meV & -2.1 meV & -4.3 meV & -0.1 meV\\
& K-L$_1$(c) & -3650.6 meV & -2.6 meV & -5.5 meV & -1.4 meV \\
& K-L$_2$(c) & -3650.6 meV & -2.6 meV & -5.5 meV & -1.4 meV \\
\hline
\end{tabular}
\end{ruledtabular}
\caption{\label{tab:table2} The same as in \autoref{tab:table1A}, however for \it{ H-type} structures.}
\end{table*}

\begin{table*}
\begin{ruledtabular}
\begin{tabular}{c|cccc}
\multicolumn{2}{c}{} & $\mathcal{E}^{\lambda}_{l}$ & $\alpha^{\lambda}_l$ & $\beta^{\lambda}_{l}$ \\
\hline
\hline
\multirow{4}{*}{$\text{MoS}_2-\text{WS}_2$} & Mo(v) & -6141.9 meV & -4.944 meV & -7.400 meV\\
& W(v) & -5951.6 meV & -2.000 meV & -7.589 meV\\
& Mo(c) & -4357.4 meV & -4.944 meV & -7.511 meV\\
& W(c) & -4005.2 meV & -1.900 meV & -7.700 meV\\

\hline
\multirow{4}{*}{$\text{MoSe}_2-\text{WSe}_2$} & Mo(v) & -5537.3 meV & -4.156 meV & -5.889 meV\\
& W(v) & -5361.6 meV & -1.467 meV & -5.856 meV\\
& Mo(c) & -3977.5 meV & -4.389 meV & -6.178 meV\\
& W(c) & -3668.3 meV & -1.500 meV & -5.978 meV\\

\hline
\multirow{4}{*}{$\text{MoS}_2-\text{MoS}_2$} & L$_1$(v) & -6158.8 meV & -3.456 meV & -7.467 meV\\
& L$_2$(v) & -6158.8 meV & -3.456 meV & -7.467 meV\\
& L$_1$(c) & -4374.2 meV & -3.444 meV & -7.589 meV\\
& L$_2$(c) & -4374.2 meV & -3.444 meV & -7.589 meV\\

\hline
\multirow{4}{*}{$\text{MoSe}_2-\text{MoSe}_2$} & L$_1$(v) & -5556.6 meV & -2.856 meV & -6.044 meV\\
& L$_2$(v) & -5556.6 meV & -2.856 meV & -6.044 meV\\
& L$_1$(c) & -3997.1 meV & -2.978 meV & -6.244 meV\\
& L$_2$(c) & -3997.1 meV & -2.978 meV & -6.244 meV\\

\hline
\multirow{4}{*}{$\text{WS}_2-\text{WS}_2$} & L$_1$(v) & -5935.2 meV & -3.433 meV & -6.422 meV\\
& L$_2$(v) & -5935.2 meV & -3.433 meV & -6.422 meV\\
& L$_1$(c) & -3996.2 meV & -3.567 meV & -6.456 meV\\
& L$_2$(c) & -3996.2 meV & -3.567 meV & -6.456 meV\\

\hline
\multirow{4}{*}{$\text{WSe}_2-\text{WSe}_2$} & L$_1$(v) & -5344.9 meV & -2.778 meV & -6.067 meV\\
& L$_2$(v) & -5344.9 meV & -2.778 meV & -6.067 meV\\
& L$_1$(c) & -3655.4 meV & -2.800 meV & -6.220 meV\\
& L$_2$(c) & -3655.4 meV & -2.800 meV & -6.220 meV\\

\hline
\end{tabular}
\end{ruledtabular}
\caption{\label{tab:table3} Fit parameters from \autoref{eq:fitting}) determining the stacking-dependent band edge energies for R-type structures.}
\end{table*}

\begin{table*}
\begin{ruledtabular}
\begin{tabular}{c|cccc}
\multicolumn{2}{c}{} & $\mathcal{E}^{\lambda}_{l}$ & $\alpha^{\lambda}_l$ & $\beta^{\lambda}_{l}$ \\
\hline
\hline
\multirow{4}{*}{$\text{MoS}_2-\text{WS}_2$} & Mo(v) & -6142.8 meV & -0.367 meV & -1.144 meV\\
& W(v) & -5950.0 meV & -1.333 meV & -1.411 meV\\
& Mo(c) & -4357.8 meV & -0.556 meV & -1.367 meV\\
& W(c) & -4003.2 meV & -1.344 meV & -1.278 meV\\

\hline
\multirow{4}{*}{$\text{MoSe}_2-\text{WSe}_2$} & Mo(v) & -5538.7 meV & -0.389 meV & -1.044 meV\\
& W(v) & -5362.8 meV & -1.611 meV & -1.189 meV\\
& Mo(c) & -3978.7 meV & -0.622 meV & -1.389 meV\\
& W(c) & -3669.8 meV & -1.611 meV & -0.933 meV\\

\hline
\multirow{4}{*}{$\text{MoS}_2-\text{MoS}_2$} & L$_1$(v) & -6158.5 meV & -0.378 meV & -0.122 meV\\
& L$_2$(v) & -6158.5 meV & -0.378 meV & -0.122 meV\\
& L$_1$(c) & -4373.7 meV & -0.656 meV & -0.511 meV\\
& L$_2$(c) & -4373.7 meV & -0.656 meV & -0.511 meV\\

\hline
\multirow{4}{*}{$\text{MoSe}_2-\text{MoSe}_2$} & L$_1$(v) & -5557.1 meV & -0.500 meV & -0.156 meV\\
& L$_2$(v) & -5557.1 meV & -0.500 meV & -0.156 meV\\
& L$_1$(c) & -3997.1 meV & -0.578 meV & -0.144 meV\\
& L$_2$(c) & -3997.1 meV & -0.578 meV & -0.144 meV\\

\hline
\multirow{4}{*}{$\text{WS}_2-\text{WS}_2$} & L$_1$(v) & -5936.6 meV & -0.344 meV & -0.033 meV\\
& L$_2$(v) & -5936.6 meV & -0.344 meV & -0.033 meV\\
& L$_1$(c) & -3997.0 meV & -0.767 meV & -0.367 meV\\
& L$_2$(c) & -3997.0 meV & -0.767 meV & -0.367 meV\\

\hline
\multirow{4}{*}{$\text{WSe}_2-\text{WSe}_2$} & L$_1$(v) & -5345.0 meV & -0.489 meV & -0.144 meV\\
& L$_2$(v) & -5345.0 meV & -0.489 meV & -0.144 meV\\
& L$_1$(c) & -3655.1 meV & -0.555 meV & -0.067 meV\\
& L$_2$(c) & -3655.1 meV & -0.555 meV & -0.067 meV\\

\hline
\end{tabular}
\end{ruledtabular}
\caption{\label{tab:table4} The same as in \autoref{tab:table3}, however for H-type structures.}
\end{table*}

\begin{table*}
\begin{ruledtabular}
\begin{tabular}{cccccccc}
& & $R_h^h$ & $R_h^M$ & $R_h^X$ & $H_h^h$ & $H_h^X$ & $H_h^M$ \\
\hline
\hline
\multirow{4}{*}{$\text{MoS}_2-\text{WS}_2$} & K$_v(^{\prime})$ & 15.5 meV & 0 & 0 & 51.1 meV & 0 & 0\\
& K$_c(^{\prime})$ & 7.5 meV & 0 & 0 & 0 & 12.5 meV & 0\\
& $\Lambda_c$ & 124.7 meV & 175.6 meV & 160.6 meV & 206.7 meV & 152.7 meV & 119.8 meV\\
& $\Gamma_v$  & 215.9 meV & 387.8 meV & 392.0 meV & 373.4 meV & 344.6 meV & 230.8 meV\\
\hline
\multirow{4}{*}{$\text{MoSe}_2-\text{WSe}_2$} & K$_v(^{\prime})$ & 19.4 meV & 0 & 0 & 57.8 meV & 0 & 0\\
& K$_c(^{\prime})$ & 9.9 meV & 0 & 0 & 0 & 19.1 meV & 0\\
& $\Lambda_c$ & 139.6 meV & 186.7 meV & 172.6 meV & 212.1 meV & 165.1 meV & 138.8 meV\\
& $\Gamma_v$  & 200.4 meV & 365.2 meV & 367.3 meV & 349.0 meV & 331.0 meV & 219.0 meV\\
\hline
\multirow{4}{*}{$\text{MoS}_2-\text{MoS}_2$} & K$_v(^{\prime})$ & 15.4 meV& 0 & 0 & 44.9 meV & 0 & 0\\
& K$_c(^{\prime})$ & 2.3 meV & 0 & 0 & 0 & 6.1 meV & 0\\
& $\Lambda_c$ & 131.5 meV & 184.2 meV & 184.2 meV & 209.4 meV & 169.6 meV & 137.0 meV\\
& $\Gamma_v$  & 208.3 meV & 391.0 meV & 391.0 meV & 352.9 meV & 352.7 meV & 242.5 meV\\
\hline
\multirow{4}{*}{$\text{MoSe}_2-\text{MoSe}_2$} & K$_v(^{\prime})$ & 19.3 meV& 0 & 0 & 56.3 meV & 0 & 0\\
& K$_c(^{\prime})$ & 5.3 meV & 0 & 0 & 0 & 11.9 meV & 0\\
& $\Lambda_c$ & 147.1 meV & 194.7 meV & 194.7 meV & 224.8 meV & 177.8 meV & 145.0 meV\\
& $\Gamma_v$  & 198.4 meV & 365.8 meV & 365.8 meV & 357.4 meV & 327.6 meV & 208.3 meV\\
\hline
\multirow{4}{*}{$\text{WS}_2-\text{WS}_2$} & K$_v(^{\prime})$ & 21.1 meV& 0 & 0 & 55.1 meV & 0 & 0\\
& K$_c(^{\prime})$ & 0.2 meV & 0 & 0 & 0 & 0.8 meV & 0\\
& $\Lambda_c$ & 152.7 meV & 190.7 meV & 190.7 meV & 227.6 meV & 180.6 meV & 140.9 meV\\
& $\Gamma_v$  & 235.7 meV & 362.2 meV & 362.2 meV & 359.5 meV & 357.3 meV & 232.1 meV\\
\hline
\multirow{4}{*}{$\text{WSe}_2-\text{WSe}_2$} & K$_v(^{\prime})$ & 23.1 meV & 0 & 0 & 66.9 meV & 0 & 0\\
& K$_c(^{\prime})$ & 0.6 meV & 0 & 0 & 0 & 1.5 meV & 0\\
& $\Lambda_c$ & 155.2 meV & 202.8 meV & 202.8 meV & 236.6 meV & 184.0 meV & 151.6 meV\\
& $\Gamma_v$  & 201.7 meV & 356.0 meV & 356.0 meV & 346.9 meV & 321.3 meV & 208.9 meV\\
\end{tabular}
\end{ruledtabular}
\caption{\label{tab:table5} Tunnelling strength as extrapolated from first-principle calculations for  different homo- and heterobilayers.}
\end{table*}

\providecommand{\noopsort}[1]{}\providecommand{\singleletter}[1]{#1}%

\end{document}